\documentclass{article}
\usepackage{aas2pp4}
\usepackage{flushrt}

\def\Msun{M_{\odot}\ }

\def\kms{\hbox{km~s}$^{-1}$}

\def\M10{{\times 10^{10} M_{\odot}\ }}

\begin{document}

\title{Self-regulating galaxy formation as an explanation for the
Tully-Fisher relation}

\author{D. Elizondo, G. Yepes}
\affil{Departamento de F\'{\i}sica Te{\'o}rica C-XI, Universidad
Aut{\'o}noma de Madrid, Cantoblanco 28049, Madrid, Spain}

\author{R. Kates, V. M{\"u}ller}

\affil{Astrophysikalisches Institut Potsdam, Potsdam, Germany}

\and
\author{A. Klypin}
\affil{Department of Astronomy, New Mexico State University,
 Las Cruces, NM 88001}



\begin{abstract} Using 3D hydrodynamical simulations of galaxy
formation with supernova feedback and a multiphase medium, we
derive theoretical relations analogous to the observed
Tully-Fisher (TF) relations in various photometric bands.  This
paper examines the influence of self-regulation mechanisms
including supernova feedback on galaxy luminosities
and the TF relation in three
cosmological scenarios (CDM, $\Lambda $CDM and BSI (broken scale
invariance)). Technical questions such as dependence on
resolution, galaxy finding algorithms, assignment procedure for
circular velocity are critically examined.

The luminosity functions in the $B$ and $K$ bands are quite
sensitive to supernova feedback at the faint end studied here.  
We find that the faint end of the $B$-band luminosity function 
($-18 \leq M_B \leq -15$) is $\alpha \approx $
 -(1.5 to 1.9). This slope is steeper than
the Stromlo-APM estimate, but in rough agreement 
with the recent ESO Slice Project estimates.

The galaxy catalogs derived from our hydrodynamical simulations lead
to an acceptably small scatter in the theoretical TF relation
amounting to $\Delta M =0.2-0.4$ in the $I$ band, and increasing
by 0.1 magnitude from the $I$-band to the $B$-band.  Our results
give strong evidence that the tightness of the TF relation cannot
be attributed to supernova feedback alone.  However, although
eliminating supernova feedback hardly affects the scatter, it
does influence the slope of the TF relation quite sensitively. 
With supernova feedback, $L\propto V_c^{3-3.5}$ 
(depending on the strength of supernova feedback). Without
it, $L\propto V_c^{2}$ as predicted by the virial
theorem with constant $M/L$.

The TF relation reflects the complex connection between depths
of galaxy potential wells and the supply of gas for star
formation.  Hydrodynamic simulations provide direct information on
this connection and its dependence on modeling parameters.
Because of the small number of phenomenological
parameters in our approach, it can serve as a useful laboratory
for testing various hypotheses and gaining insight into the
physics responsible for the scatter, slope, and
amplitude of the TF relation.

\end{abstract}



\section{Introduction}

The tight observed correlation between luminosity $(L)$ and rotational velocity
$(V)$ of spiral galaxies, $L\propto V^{3.5-4}$,  (Tully \& Fisher, 1977;
Pierce \& Tully, 1988, 1992; hereafter TF relation) has proven to be 
one of the most useful tools in cosmology, most notably
when applied to modeling of large-scale velocity
fields and to the determination of the Hubble constant
(see, e.g., Strauss \&  Willick, 1995). However, the TF relation is essentially
empirical and has not yet received a comprehensive 
physical explanation within our current attempts to understand
the mechanisms of galaxy formation and evolution. 
Although some relationship between luminosity and rotational velocity
would certainly be expected from the virial theorem ($v^2 \propto GM/R$),  
the slope of the relation cannot be explained by a simply 
combining the virial theorem 
with a constant mass to luminosity assumption.  Processes determining 
the luminosity of dark-matter halos must be an essential ingredient 
in any explanatory model. Hence, a key  issue  in understanding  
the origin of the TF relation  is the effectiveness of star formation and
its back-reaction on the galaxy formation process. 
A reliable theory of galaxy formation and evolution should
be capable of answering the following questions:
(i) why the relation has  a small scatter;
(ii)  why the slopes and zero points are as observed;
(iii) what additional variables (either potentially
observable or hidden) could explain the residuals.

Different galaxies have different merging histories and different
gas accretion rates. These processes affect star formation
rates and hence could introduce 
scatter into the distribution of galaxy luminosities and colors. 
Thus the first
question (low observed scatter) already represents a severe challenge
for theory.  Two mechanisms for explaining the low scatter were investigated by
Eisenstein \& Loeb (1996).  First, one could suppose that galaxies
formed prior to $z=1$ and accreted only small amounts of material 
subsequently. However, neither the standard
hierarchical clustering picture nor the particular simulations 
reported below support this hypothesis.  Secondly, a
strong feedback process could decouple the overall luminosity of spiral
galaxies from details of their formation histories.
This kind of mechanism is quite plausible, because it is known that dynamical
systems with nonlinear feedback can be efficient at ``forgetting''
initial conditions and at reducing 
sensitivity to some parameters.

The second question (zero point and steepness of the slope) involves details of
the star formation history within developing galaxies.  Therefore, any realistic
model of galaxy formation intending to answer this question needs to take into
account at least the main physical mechanisms influencing the rate of star
formation:  gas dynamics and radiative processes.

Finally, the processes responsible for the explanation of the TF-relation
involve variables (such as the merging and accretion history of the galaxy) that
do not appear explicitly in the relation and which in this sense are ``hidden''
to the observer.  A theoretical understanding, besides its importance as a basic
scientific achievement, could offer several important practical
advantages compared to a
purely empirically-based relation:  First, corrections to the relation on the
basis of available observable properties could improve the accuracy of distance
determinations.  Second, the existence of possible systematic biases (such as
hidden dependence on environment) is controversial (Pierce \& Tully, 1988, 1992,
Elizondo et al, 1998).  If present, environmental bias in distance determination
could have important consequences for mapping the large-scale velocity field.
Third, if the form of the TF relation is sensitive to the model of structure
formation, the relation could also be useful in discriminating among
cosmological scenarios.

The analysis of Eisenstein and Loeb (1996) shows that the low scatter of the TF
relation is unlikely to have resulted from initial conditions.  This conclusion
suggests that some regulation or feedback mechanism(s) must be at work.
Whatever mechanism operates, it must keep both the mass-luminosity relation and
the mass-velocity relation tight.  Nonlinear processes that introduce feedback
into galaxy formation include energy input and metal enrichment due to supernova
explosions as well as hydrodynamics.  All of these processes would be expected
to contribute significantly to self-regulation of galaxy luminosities.

\begin {itemize} 
\item Energy input by supernovae:  The theory of the interstellar medium (McKee
\& Ostriker, 1977) implies that much of the energy of supernovae is converted to
evaporation of cold clouds and thus has an important influence on subsequent
star formation.  The relative fraction of the energy used to evaporate clouds
characterizes the degree of supernova feedback, as discussed below. In the
context of galaxy formation and evolution, the feedback associated with
injection of heat into the gas due to supernovae was studied by Yepes, Kates,
Khokhlov \& Klypin (1997; YK$^3$) and found to have a significant
self-regulating effect on the star formation history.

\item Metal enrichment: During supernova explosions, 
metal enriched gas is distributed
over large volumes within a galaxy and may even be transported to
the intergalactic medium, in the case of small halos. 
Metals strongly increase the subsequent cooling
rate, resulting in an increased star formation rate. 
Observationally, metallicity
gradients in galaxies provide evidence for 
the effectiveness of this mechanism in the central regions of galaxies.

\item Hydrodynamic feedback: The transport of gas depends on pressure 
gradients
and thus it is particularly sensitive to the injection of large amounts 
of energy into the gas. Hence, in addition to the 
self-regulation due to cloud evaporation, there is also a strong
self-regulating tendency due to gas flowing out of galaxies into 
lower-density regions, where the cooling time scale is longer.
Gas which remains outside the galaxy will not contribute to 
further star formation inside the galaxy, and star formation
from the gas fraction that does eventually fall back in will be delayed. 

\end {itemize}
All those feedback mechanisms are tightly 
coupled, and it remains to be seen how such inherently complex 
processes can interact to yield relatively simple characteristics and
relationships such as the TF relation.

The TF relation has been investigated using so-called ``semi-analytical'' models
(Kauffmann et al., 1993 (KWG); Cole et al., 1994 (CAFNZ); Heyl et al., 1995
(HCFN), Avila-Reese, Firmani \& Hern{\'a}ndez 1998).  Parametrising the merging
history, gas cooling and star formation histories allows efficient generation of
large synthetic galaxy data samples.  The parameter space can then be
investigated by testing the agreement with locally observed astrophysical
processes.  For example, semi-analytical models have been used for studying the
influence of different cosmological evolution rates on galaxy formation
processes. As a first step in understanding the processes involved in the
TF relation, these investigations provide quite a bit of useful information. 

As in the $YK^3$ model used here, the baryonic content of a halo in
semi-analytic models (e.g.  CAFNZ) typically consists of three components:  hot
gas, cold star-forming clouds, and stars.  However, because gas dynamics are not
explicitly simulated, the semi-analytical models require various uncontrolled
assumptions such as specification of the shape of a halo.  Generally, the models
idealize a galaxy halo as a round object with an assumed density profile
(typically an isothermal profile $\rho\propto r^{-2}$) and constant gas
temperature (equal to the virial temperature).  An assumed density profile is of
course not necessarily even close to a solution of the gas dynamical equations.
For example, during large mergers, it is neither possible to fit the density by
a simple spherical density profile, nor may the gas be treated as isothermal.
In particular, an understanding of environmental influence on galaxy properties
(Elizondo et al., 1998) 
requires a hydrodynamic treatment, because gas flows are neither
spherically symmetric nor confined to the interior of halos. 

Since dynamical modeling is not included, semi-analytic models do not provide
the flexibility of hydrodynamic simulations, which they often try to compensate
for by involving extra fitting parameters.  For example, as reported by HCFN,
none of the models studied was able to reproduce the observed galaxy luminosity
function and the TF relation simultaneously, even allowing the mass-to-light
ratio to be a free parameter.  It is possible to reconcile some of the
discrepancies by neglecting star formation in halos with circular velocities
greater than 500 \kms, as done by KWG, but this prescription was criticized by
HCFN because it is preferable to avoid introducing ad hoc assumptions.  HCFN
also discussed the uncertainties associated with the identification of
rotational velocities of galaxies with those of halos.  Note that in the
semi-analytical approaches, the gas is assumed isothermal out to the virial
radius (radius of a sphere of overdensity 200), which as pointed out by HCFN may
be violated.  The scatter in the TF relation was reported by CAFNZ to be about
0.5 mag at 200 \kms, similar to the observed values, but with an increase in the
scatter for low-velocity halos.  

Hydrodynamical simulations of galaxy formation (e.g.  Katz 1992; Cen \& Ostriker
1992; Navarro \& White 1994; Steinmetz \& M{\"u}ller 1994, Yepes et al. 1995,
YK$^3$) offer the clear advantage of actually solving the gas dynamical
equations.  Hence, they can model the dynamical behavior of both the gas and the
dark matter component at some level of accuracy in all conceivable situations,
even when large mergers occur.  Nevertheless, they also have their own problems,
such as limited resolution.  This problem should gradually decrease in severity
as computational costs fall.  A more serious problem is that physical processes
occurring far below the resolution limit, such as star-gas interactions, can
have important effects and must therefore be included by hand.  However, despite
these problems, hydrodynamical approaches are a complementary tool to
investigate the complex area of galaxy formation and to gain insight into the
origin of the properties of galaxies we observe.

The structure of this paper is as follows:  we start with a brief review of 
basic methodology we have used for studying the
theoretical TF relation, including  the model, simulation techniques, and 
galaxy identification schemes. Next we describe and critically discuss the 
assignment of rotational velocity to the {\em numerical} galaxies 
({\S} \ref{sec:vc}).  The 
fits to the luminosity-rotational velocity relation in the different color bands
and their dependence on supernovae feedback are described in {\S}
\ref{sec:fit}.  Section \ref{sec:resolution} is devoted to the study of the
influence of numerical resolution and the galaxy finding algorithm on our
results.  In {\S} \ref{sec:luminosity} we show our estimates of the faint-end of
luminosity functions in B and K bands and compare them with observational data.
Finally, Section \ref{sec:discussion} is devoted to the discussion of results
and conclusions.

\begin{table*}[t]
\caption{Parameters of the scenarios \label{tab:para1}}
\small
\begin{tabular}{lccc}
  \hline Parameters & CDM  & $\Lambda$CDM  & BSI
   \\ \hline h & 0.5 & 0.7 & 0.5 \\
  $\Omega_{\Lambda}$ & 0 & 0.65 & 0 \\ $\Omega_{B}$
  & 0.051 & 0.026 & 0.051 \\ 
  $\Omega_{dm}$ & 0.949 & 0.324 & 0.949 \\ 
  rms fluctuation for 8$h^{-1}$Mpc ($\sigma_8$) & 1.2 &1.0 & 0.6\\
   $\Omega_{dm}$ & 0.949 & 0.324 & 0.949 \\ 
  Simulation Box at $z=0$, (Mpc) & 5 & 5 & 5 \\ 
  Number of particles & $128^3$ &  $128^3$ (256$^3$) & $128^3$ \\ 
  Cell size at $z=0$, (kpc) & 39 & 39 (19.5)& 39 \\
 Mass  Resolution for dark matter ($10^6 M_{\odot}$) & 4 & 2.7 (0.3) & 4 \\ 
 \hline({\small In brackets: parameters of the high-resolution simulation}) & & & 
\end{tabular}
\end{table*}

\section{Methods}
\label{sec:methods}
\subsection{Physical processes}
\label{subsec:procs}

To study the origin of the TF relation, we have used 3D gas dynamical
simulations which include models for the most important physical processes in
galaxy formation, as described in $YK^3$:  these include gas dynamics, radiative
and Compton cooling, star formation, star-gas interactions in the form of a
two-phase approximation for the interstellar medium, and supernovae feedback. 
The universe is modeled as a 4-component medium consisting
of dark matter, stars, ``hot'' or ambient gas, and cold clouds.  
Star formation is modeled by converting cold gas
at a certain rate into discrete star particles. Each such event may
be idealized as representing a small ``starburst.''  Stellar population synthesis
models (Bruzual \& Charlot, 1993) are used to derive the luminosities and colors
of the galaxies as a superposition of the starburst contributions.

The equations for the multifluid dynamics including star formation
and supernova feedback, effects of resolution, and the values
of ``chemical'' parameters were reported in detail in YK$^3$. 
Here we summarize the physical model in a slightly simplified form.

In each 
volume element, the amount of cold gas
$m_{\rm cold}$ capable of producing stars is regulated by the mass of the hot
gas $m_{\rm hot}$ that can cool on the time scale $t_{\rm cool}$, 
by the rate of forming new stable stars, and by the effects
of supernova formation, 
which reheat and evaporate
cold gas.  The supernovae formation rate is assumed to be proportional to the
mass of cold gas:  
$\dot m_{\rm SN}=\beta m_{\rm cold}/t_*\equiv m_{\rm
SN}/t_*$, where $t_*=10^8$ yr is the time scale for star formation, and 
$\beta$ is the fraction of mass of newly formed stars that explode as supernovae
($\beta=0.12$ for the Salpeter IMF).  Each $1\Msun$ of supernovae dumps
$E_{SN} = 4.5\times 10^{49}$ ergs of heat into the interstellar medium
 {\it and} evaporates a mass $A\cdot\Msun$ of cold gas.   The constant
 $A$ will be referred to below as the ``supernova feedback parameter'' and
is the most important parameter in the model.  
 
The mass transfer between different components satisfies
\begin{eqnarray}
\dot m_{\rm hot}& =& {Am_{\rm SN}\over t_*}-{m_{\rm hot}\over t_{\rm
cool}},\nonumber \\
\dot m_{\rm cold}& =&-{(m_{\rm cold} -m_{\rm SN})\over t_*}-\dot m_{\rm
hot},\nonumber \\ 
t_{\rm cool}&= &{kT_{\rm hot}\mu m_H\over \rho_{\rm hot}\Lambda(T_{\rm
hot})C(\gamma-1)},
\end{eqnarray}
where $k$ is the Boltzmann constant, $T_{hot}$ is the temperature of the 
hot gas, $\mu$ is the molecular weight per particle, $m_H$ is the
mass of hydrogen, $\rho_{\rm hot} $ is the density of the hot gas,
$\Lambda$ is the cooling rate, and $\gamma = 5/3$ is the ratio of
specific heats. The adjustable parameter 
$C=1-10$ in the cooling time mimics the effects of
unresolved substructure. 

In the regime of active
star formation, the main effects defining the temperature of the gas
are supernova feedback and evaporation of the cold gas, which
absorbs a significant fraction of the supernova energy:
\begin{equation}
m_{\rm hot}\dot T_{\rm hot}={m_{\rm SN}\over t_*}[T_{\rm SN}-A(T_{\rm
hot}-T_{\rm cold})].
\end{equation}
In this equation, $T_{\rm cold}\sim 10^4$K is the temperature of the
cold gas, and $T_{\rm SN}=1.1\times 10^8$K is a measure of the energy
released by a supernova: $kT_{\rm SN}=(\gamma-1)\mu m_H\langle E_{\rm
SN}\rangle/\langle m_{\rm SN}\rangle$. In order to mimic the effects
of photoionization by quasars and AGNs, gas with overdensity less
than 2 was kept at a constant temperature of 
$10^{3.8}$K (Giroux \& Shapiro, 1986; Petitjean, M{\"u}cket, \& Kates,
1995).  YK$^3$ found
that results are relatively insensitive to the cooling enhancement
factor $C$. For example, changing the factor from 1 to 10 gives
practically the same results. This occurs because of
the rapid increase in gas density after gravitational collapse sets in, 
leading to very effective cooling even for $C=1$.

\subsection{Overview of numerical simulations}
\label{subsec:numsim}

The $YK^3$ code used here employs
a combined N-body (PM) and hydrodynamic technique.  The
Eulerian hydrodynamical equations in the expanding Universe are solved on a
grid using the {\em Piecewise Parabolic Method\/} (Colella and
Woodward, 1984). The method includes
high-resolution shock capturing technique and is
capable of resolving hydrodynamic shocks without artificial viscosity.

In what follows, all quantities
are expressed in physical units in which the Hubble constant of the
corresponding cosmological scenario has been included (see Table
\ref{tab:para1}).

A set of 11 simulations were first performed and 
analyzed for each of the cosmological
scenarios used.  This provides sufficiently large catalogs of ``numerical
galaxies'' to allow statistically significant comparisons with observational
quantities.  These simulations were run using $128^3$ particles on a $128^3$
grid, both for the dark matter and the gas dynamics.  We chose a box size of 5
Mpc as a compromise between the requirement of taking a representative part of
the universe and having enough resolution to describe the mean star formation
activity in a cell of the simulation volume.  The cell size is thus
39 kpc, and the mass per particle is $\sim 4 \times 10^6 M_{\odot}$. 
The supernova feedback parameter in this simulation set was $A=200$.


To study the effects of numerical resolution, we ran a test
simulation for $\Lambda$CDM 
with the same box size and chemical parameters as the 
11  realizations,  but with $256^3$ particles and cells 
(i.e. 19.5 kpc comoving and $3.3\times 10^5 M_\odot$
mass per particle). We then reran this simulation  at  lower
resolution ($128^3$ particles and cells). The initial particle
distributions for all simulations were set up by the Zeldovich 
approximation. The displacement field used for the  $128^3$ grid was
generated by a numerical average  over the eight nearest cells of 
the displacement field in the $256^3$ grid. This procedure allowed us
to identify the same  (massive)  halos at the end of evolution in both
simulations which made possible a reliable comparison of the effects of
resolution in the final observational properties of the halo
distribution. Increasing the resolution by a factor of two
did not result in significant changes in global parameters (mass,
luminosity) of bright galaxies. 
In particular no systematic change in the TF fit was found. 
Details and additional tests are reported in {\S} \ref{sec:resolution}.

\begin{table*}[t]
\caption{Number of galaxies in the simulations. \label{tab:para2}}
\small
\begin{tabular}{lcccc}
  \hline  Model & Feedback & \# realizations & \# galaxies &
\# bright galaxies   ($M_B < -16$)  \\ 
  & Parameter &  &  &  \\ \hline

  & A=200 & 11 & 447 & 73 \\ 
 CDM & A=50 & 6 & 262 & 34 \\ 
 & A=0 & 6 & 288 & 77 \\  \\ 
  & A=200 & 11 & 442 & 91 \\ 
 $\Lambda$CDM & A=50 & 6 & 260 & 33 \\ 
 & A=0 & 6 & 285 & 94 \\ \\ 
 & A=200 & 11 & 649& 144 \\ 
 BSI & A=50 & 6 & 318 & 36 \\ 
 & A=0  & 6 &317 & 182 \\ 
  \hline
\end{tabular}
\end{table*}

To study the effects of supernovae feedback on the final observational
properties of simulated galaxies, we repeated 6 of the 11 simulations for
each cosmological model with the feedback parameter values $A=50$
(strong gas  reheating) and $A=0$ (no  reheating or mass
transfer).  Note, however, that we still
keep  metallicity enrichment in this case (see Table
\ref{tab:para2}). A detailed discussion of  the effects of
supernova feedback on the TF relation is given in  {\S} \ref{sec:fit}.

The numerical code is parallelized to a degree of 95\% for 
symmetric-multiprocessor computer architectures.
All  simulations reported here 
were performed in SGI Power Challenge and Origin 2000 supercomputers 
at the European Center for Parallelism in Barcelona (CEPBA).

\subsection{Cosmological models}

The parameters of the cosmological scenarios studied here are summarized in
Table 1:  We consider a standard CDM scenario, a low-density $\Lambda$CDM
scenario, and a broken-scale-invariant (BSI) scenario which is based on a double
inflation model.    The BSI scenario
was analyzed by Kates et al.  (1995) using a particle-mesh code that included a
thermodynamic model for the gas.  The post-recombination power spectrum of the
BSI model (for an analytic fit, see Kates et al., 1995) is similar to that of
the $\tau$CDM scenario, in which 
the dark matter consists of a decaying neutrino.


\begin{figure}[htbp]
  \begin{center}
    \leavevmode
\plotone{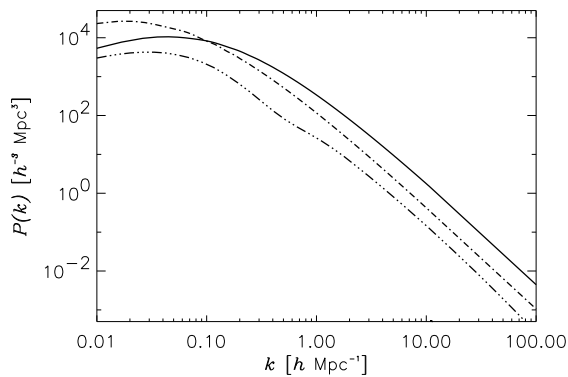}
    \figcaption[TULLY1.ps]{Initial power spectra for the CDM (solid), 
    $\Lambda$CDM (dashed) and BSI (dash-dotted) models.
    \label{fig:power}}
  \end{center}
\end{figure}

Within the simulated range of scales, the
perturbation spectra for all three scenarios follow a power law, $P(k) \propto
k^{-2}$.  The power spectra for the three models are shown in Fig.  \ref{fig:power}.
Although all three spectra are normalized to COBE, the spectral amplitudes
differ considerably in our 5 Mpc box, because the normalization is performed on
a scale much larger than the box.  For example, the amplitude of the initial
perturbations for BSI is about one-third that of CDM.  That is, our BSI
simulation can be considered as a CDM simulation with an effective normalization
corresponding to $\sigma_8=0.35$.  Hence, any inferences to be drawn should not
be seen as statements about the viability of scenarios, but rather as a source
of information on how the merging rate (CDM versus $\Lambda$CDM) or the time
evolution (CDM versus BSI) influences the properties of galaxies.  Our
references to cosmological scenario dependence in what follows are to be understood
with this caveat in mind.


\subsection{Galaxy identification procedure}
\label{sub:gal}

Galaxy identification poses difficulties in cosmological N-body simulations.
Here we have at our disposal a substantial piece of information that is not
available in pure dark matter simulations to distinguish galaxies among dark
halos:  the star (luminosity) distribution.  In what follows, we use the term
``galaxy'' to refer to a halo with nonzero luminosity.

Our halo identification algorithm was given in YK$^3$ and is briefly reviewed
here:  We begin by looking for local dark matter density maxima with overdensity
$>$ 100 on the grid.  We then define a sphere of radius 2 cells centered at the
position of the maxima found.  The position of the center-of-mass (CM) of all
particles (dark+stars) within the sphere is then computed.  We then move the
center of the sphere to the CM position and recompute the CM of the new
particles lying within a sphere of the same radius.  This procedure is iterated
to convergence, which takes just a few iterations. In this way, we arrive to
the true local maxima of the particle distribution.  The total mass (baryons +
dark matter) within the sphere is assigned as the mass for the halo. Now, if
the mass within the sphere exceeds the mass corresponding to 
overdensity 200, we include this halo in our catalog. 
If the mass is lower, we
repeat the algorithm with a sphere of radius 1 cell.

This procedure typically gives a radius and
a mass for the halo that are well below the virial radius and close to the peak
of the rotation curve, which is the area we are interested in.  We have also
checked that with this procedure, we avoid the problem of finding halos with
substructures.  

Using the stellar population synthesis model of Bruzual \& Charlot (1993), we
assign luminosities and colors to the galaxies (halos which have formed
stars) by computing the spectral energy distribution (SED) according
to
\begin{equation}
\label{SED2}
S(\lambda,t)= \sum _{\tau_i} \Phi(\tau_i) {\cal F}(\lambda,t-\tau_i),
\end{equation}
where  $\Phi(\tau_i)$ is the mass of stars in the halo  produced at timestep
 $\tau_i$  and
${\cal F}(\lambda, t)$ is the SED due to a starburst of 1 $\Msun$
after an evolution time $t$. Convolving $S(\lambda,t)$ with the 
filter response function
$R_f(\lambda)$, we obtain the absolute luminosity $L_f(t)$ in the given band.
Combining the $L_f(t)$, we then obtain the evolution of
the theoretical color index of the galaxy.  (Of course, the color
of a galaxy that would actually be measured will be influenced
by other factors such as the interaction of starlight
with the surrounding plasma.) 

With this procedure, we typically obtain about a few tens of
galaxy-like objects  per simulation volume. In Table \ref{tab:para2} we
summarize the total number of galaxies found in all the simulations reported
here. This number of galaxies permits statements concerning trends
in the TF relation that are based on statistically significant 
estimates.

The 11 simulations carried out for each model correspond to a total 
simulated comoving volume of 1375 Mpc$^3$. Within this volume, we have
also estimated the faint end 
of the galaxy luminosity function in the $B$ and $K$ bands and to compare
these estimates with fits to the observational data, (see {\S} \ref{sec:luminosity}).

In order to test the accuracy of the galaxy identification
prescription, we have also used a somewhat more refined, but more
computationally expensive  algorithm in several of the
simulations. The procedure is basically the same as above
except that instead of
computing quantities at fixed radius (1 or 2 cells), we compute the
spherical mass and luminosity profiles. The radius is then increased
until either the overdensity in the spherical shells 
falls below 200, or the radius touches a neighboring halo. 
This procedure begins with the highest-density maxima in order to
define the large halos most reliably.  We then define the {\em optical
radius} ($r_{opt}$) of the halo, in analogy to observed
galaxies, as the radius  which  contains $\sim 80$ \% of the total
light in the $B$ band (Persic, Salucci \& Stel 1996), assuming a pure
exponential luminosity profile. The mass,
circular velocity and luminosities  for  the galaxies correspond
to the value of their respective  radial profiles at  $r_{opt}$.

The galaxies identified by these two procedures are the same, although  
their physical properties deviate slightly due to the different
limiting radius of the halos. Nonetheless, the main
results are not substantially modified. 

\section{Numerical estimates for the rotational velocity of galaxies}
\label{sec:vc}

In observed spiral galaxies, circular disk velocities are inferred from
spectroscopic linewidth estimates.  Unfortunately, this procedure does not have
a direct analog for our simulated galaxies:  First, the numerical resolution of
our simulations (39 kpc) does not allow us to
assign a morphological type by identifying a disk structure inside halos. 
Second, at the limits of resolution, circular velocities estimated
directly from the star particles in the simulations do not provide a realistic
estimate of the ``true'' circular velocity.

To overcome the first difficulty, we assign a morphological type for galaxies
based on their colors.  As it turns out, most of the galaxies in our simulations
are spirals, according to their position in the UBV color diagram (see Elizondo
et al.  1998 for a more detailed discussion).  This result is expected, because
regions with high galaxy concentrations (groups and clusters of galaxies) are
underrepresented due to the limited simulation box size.

\begin{figure}[t]
  \begin{center}
    \leavevmode
\plotone{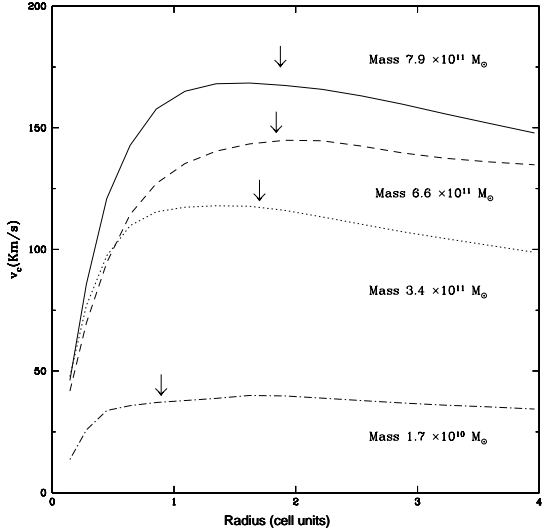}
    \figcaption[TULLY2.ps]{Circular velocity profiles 
for several halos of one of the $\Lambda$CDM
      simulations reported in the text.
 The circular velocity for  the three more massive halos is assigned at 
2 cell radii by our galaxy finding algorithm. For the less massive
halo, velocity is assigned at   1 cell radii. The arrows indicate  the
corresponding  value of $V_{2.2}$ for the different halos (see text).
    \label{fig:vcirc}}
  \end{center}
\end{figure}

To overcome the second difficulty, we have devised an operational
procedure for assigning ``rotational velocities'' to our numerical galaxies.
 We have tested this procedure for consistency in different ways.
%
We define a gravitational (or circular) 
velocity, $v_{\rm grv}=\sqrt{GM/r}$, where $M$ is the total mass
(baryons+dark matter) of a galaxy within its
assigned radius $r$  of either  1 or 2 cells.
%
To check whether our assignment of circular velocities at  discrete
radii is not introducing  significant errors into the results, 
%
%
 we have determined ``circular  velocity profiles'' for
a subsample of our numerical galaxies from their 
spherically averaged numerical density profiles. 
It turns out that $v_{\rm grv}$ nearly always coincides with the flat
part of the rotation curve, as can be seen in  Fig. \ref{fig:vcirc},
which means that our prescription for velocity assignment is reasonably
accurate.

Even for real galaxies, different observational indicators 
for rotation velocities can be defined which yield 
different estimates of the TF relation. 
In a  recent study, Courteau (1997) compares various measures of rotational
velocity from optical rotation curves and 
linewidths in spiral galaxies. He concludes that the best estimate of
the rotation velocity for TF applications, ($V_{2.2}$), is given at
the location of peak rotational velocity of a pure exponential disk
($r_{disk}=2.15h$, where in this context 
$h$ is the disk scale length; see e.g. Freeman
1970). In order to check whether in the simulations our estimate of the 
circular velocity is compatible with the value of $V_{2.2}$, we have 
indicated it with an arrow in  Fig \ref{fig:vcirc}. 
Assuming that
the luminosity inside would follow an exponential disk profile, we
can assign a characteristic disk scale length using $h=r_{opt}/3.2$.
The estimate of $r_{opt}$ is obtained by computing the radial 
luminosity profiles within the halos (see {\S} \ref{sub:gal}). 
As can be seen in the figure, the values of $V_{2.2}$ are 
very close (within 5\%) to the values of the circular velocity  
assigned by our algorithm.


\begin{figure}[htb]
\centerline{\plotone{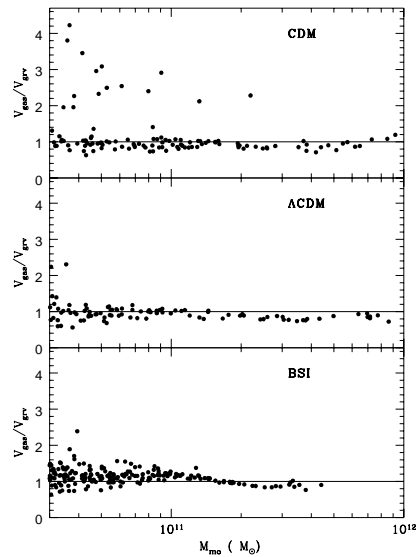}}
\figcaption[TULLY3.ps]{
The ratio of the average gas velocity
( estimated from the mass-weighted temperature) and the
circular velocity of dark halos as a function of the
mass of dark matter inside the halo ($M_{mo}$).  The points close to a
ratio of 1 represent galaxies in which the gas can be
considered as an isothermal sphere.
\label{fig:vrotvcirc}}
\end{figure}

As another test, we constructed an estimate of the rotational velocity using the
gas velocity $v_{\rm gas}$.  This estimate is based on the conjecture that the
hot gas well follows the depth of the potential wells confining the dark matter
halos and that the gas forms a rotationally supported disk.  Therefore, the
temperature of the gas should provide a good indicator for the circular velocity
of the gas.  To construct this indicator, we take the mass-weighted average of
the temperature, $<T>_M= \int \rho _{gas} T d^3x /M_{gas}$ to estimate the gas
velocity assuming an isothermal distribution, $v_{\rm gas} = \sqrt{2 k <T>_M /
\mu m_p}$ (where $m_p$ is the proton mass).  The indicators $v_{\rm grv}$ and
$v_{\rm gas}$ are compared for all three scenarios in Fig.  \ref{fig:vrotvcirc}.
The agreement of the two indicators is typically quite remarkable.  The only
outliers to the tight relation between $v_{\rm grv}$ and $v_{\rm gas}$ are
located in strongly interacting satellite galaxies, where the gas temperature is
strongly influenced by the environment, possibly involving increased star
formation and reheating due to supernova explosions.  In these cases, $v_{\rm
gas}$ would not be expected to provide a good indicator of the circular
velocity.  Most of these outliers are very faint, and hence they
do not contribute to the fits to the magnitude-circular velocity relation,
which as explained above included only galaxies with  $M_B < -16$.
Exclusion of those few outliers with $M_B< -16$ from the fits
has only a minor effect in the results.  

\begin{figure}[htbp]
  \begin{center}
    \leavevmode
\plotone{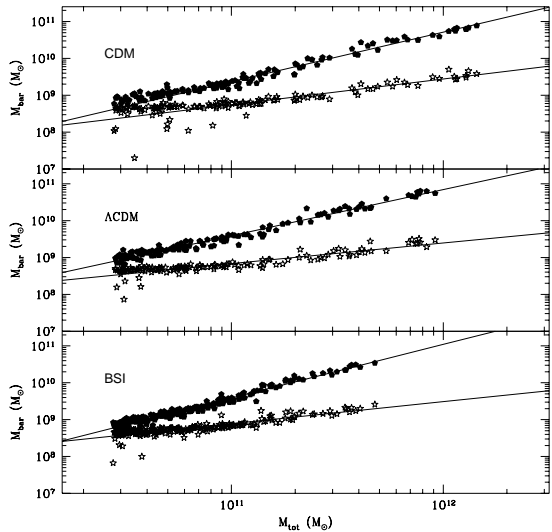}
    \figcaption[TULLY4.ps]{Relation between total mass and baryonic
      mass in the galactic halos used in the TF estimates   ($M_B <
      -16$). Solid
      points represent the total mass in baryons (stars + gas) while
      starred symbols represent the  total gas mass within the halo
      radius. Solid lines are the best fits to the points.
\label{fig:masbarfit}}
\end{center}
\end{figure}

Incidentally, most of the outliers arose in CDM
realizations.  As is plausible from the CDM spectrum (Fig.  \ref{fig:power}),
the advanced stage of evolution reached in the CDM model leads to more
interacting faint galaxies than in the other models.

We have also performed a further check on our assignment of rotation
velocities to galaxies. High-resolution hydrodynamical simulations
have found  a tight correlation between baryonic mass and
maximum rotation  velocity of the gas,
$ M_{bar} \propto V_c^{2.5}$ (Evrard, Summers \&
Davis, 1994). On the other hand, we obtain  
a very tight correlation between baryonic (gas + stars)
mass and total mass in our halos. In Fig. \ref{fig:masbarfit} we plot
this relation for the bright ($M_B\leq -16$) halos found in  
the simulations of the three
different models.  The best fits to this relation give quite similar
results for CDM and $\Lambda$CDM ($M_{bar}\propto M_{tot}^{1.3-1.25}$). 
For the less evolved BSI model, where no big galaxies 
have been formed up to the present epoch, we find a
slightly steeper slope  ($M_{bar}\propto M_{tot}^{1.45}$).  We also find a
tight correlation between the total mass and the gas content in the
halos. In this case, the correlation is pretty much the same for the three
models, $M_{gas}\propto M_{tot}^{0.6}$.  

We can now use the correlation found by Evrard, Summers and Davis (1994)
together with the correlation between total mass and baryonic mass
found in our simulations to obtain an
independent relation between total mass and 
maximum circular velocity for the halos. Doing so yields 
$M_{tot}\propto V_{c}^{1.8-1.9}$, 
which is very close to the relation we have used ($M_{tot} \propto V_c^{2}$).
Although this argument does not yield normalization constants, it provides
clear evidence that at least the slope and the scatter 
of the TF relation computed from
our simulations do not depend on the approximate estimate of the circular
velocity.

Further tests concerning the effects of numerical 
resolution and galaxy finding algorithms 
on the determination of circular velocities of galaxies 
will be discussed in {\S} \ref{sec:resolution}.


\begin{figure}[htbp]
  \begin{center}
\leavevmode
\plotone{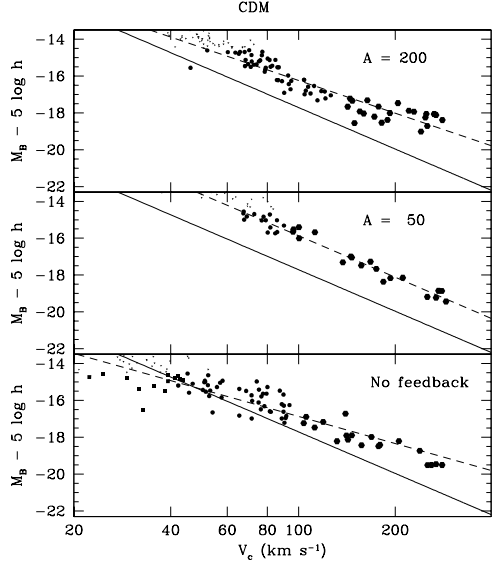}
\figcaption{(a). The TF relation for the $B$ band  for galaxies found in
  CDM simulations with different feedback parameter values.
 The upper panel ($A=200$) corresponds to simulations in which
  evaporation of cold clouds is  the most important effect. The middle
  panel, ($A=50$) correspond to simulations in which supernovae are
  very efficient at reheating the gas, but not evaporating the cold
  clouds. The lower panel  correspond to  the same simulations run
  without assuming any supernova feedback process ( $\beta =0$ see
  text.). Note, however, that we still kept metallicity enrichment  in
  this case.  The solid line shows the
  fit to the Tully-Fisher relation as
  determined by Pierce \& Tully (1992).  The dashed line
  is the best fit obtained by
  considering only simulated galaxies with $M_B <  -16$.
  Black dots represent bright galactic  galaxies  ($M_B < -16$).
  Small crosses
  represent the faint galaxies found by our   galaxy finding algorithm 
  that were not taken into account in the fit.
  \label{fig:cdmsnb}}
\end{center}
\end{figure}


\begin{figure}[htbp]
  \begin{center}
\leavevmode
\plotone{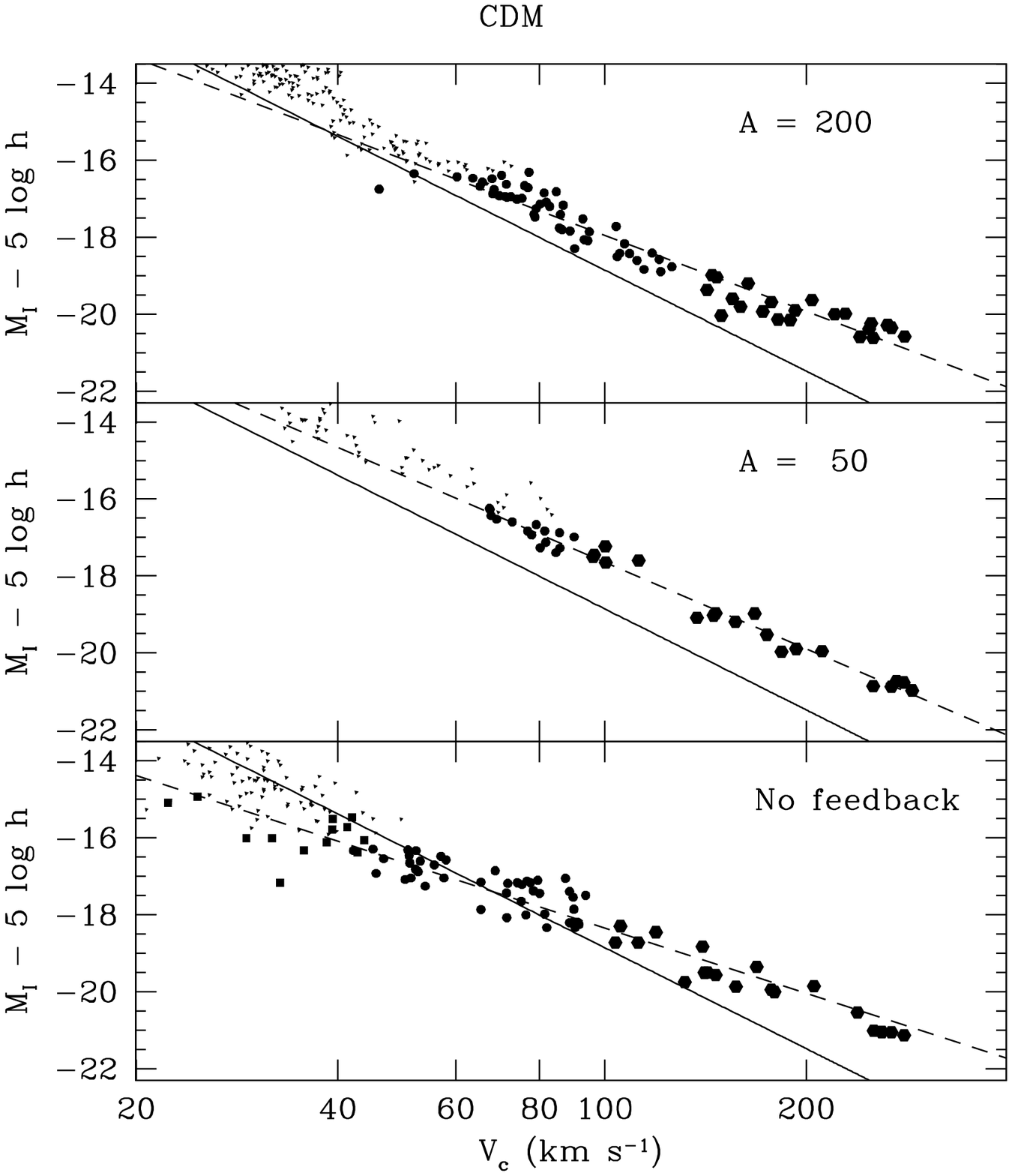}
\addtocounter{figure}{-1}
\figcaption{(b) Same as Fig \protect\ref{fig:cdmsnb}.(a)  for the $I$ band
\label{fig:cdmsni}}
\end{center}
\end{figure}


\begin{figure}[htbp]
  \begin{center}
\leavevmode
\plotone{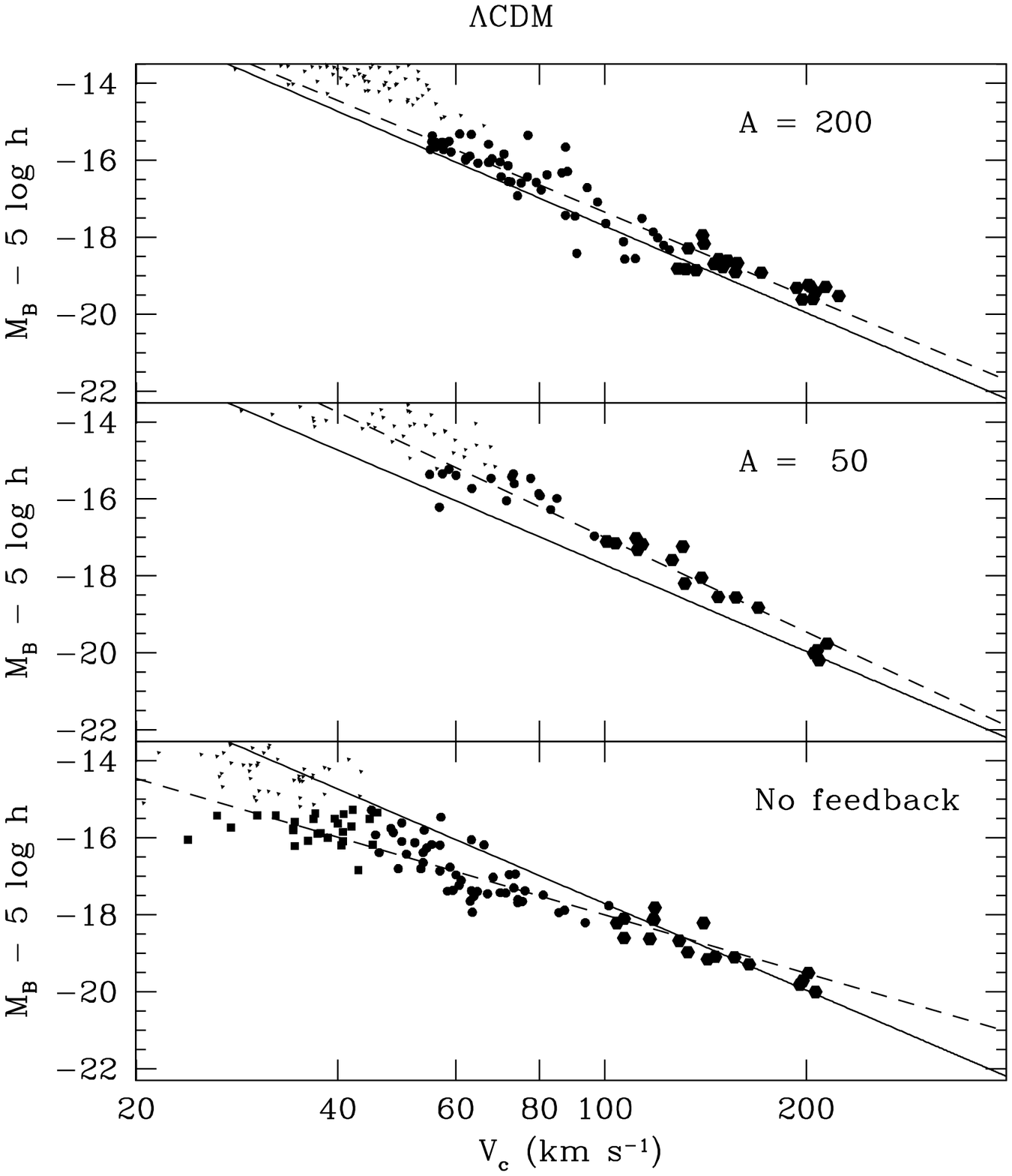}
\figcaption{(a) Same as Fig \ref{fig:cdmsnb}.(a) but for the $\Lambda $CDM model
\label{fig:lcdmsnb}}
\end{center}
\end{figure}


\begin{figure}[htbp]
  \begin{center}
\leavevmode
\plotone{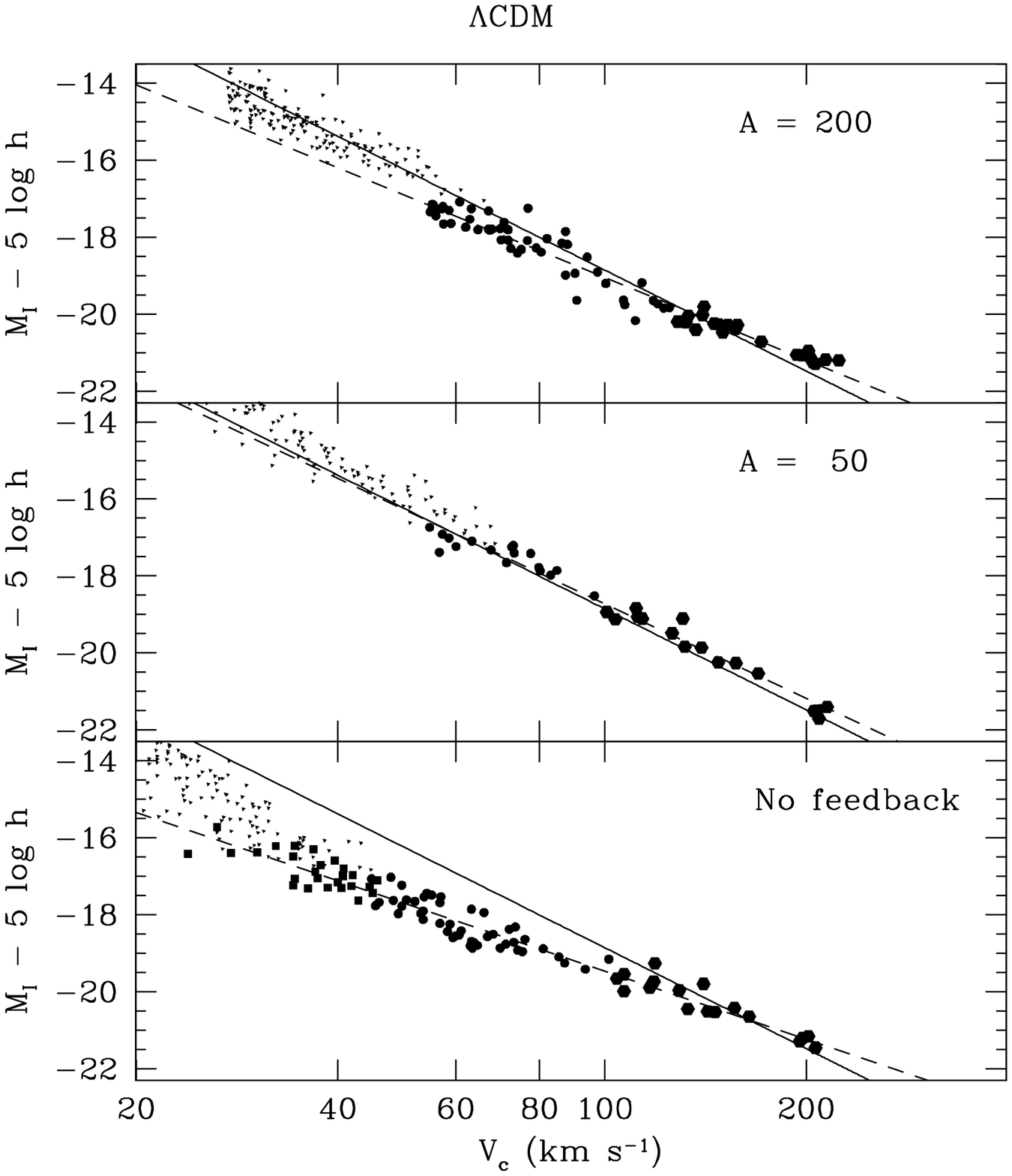}
\addtocounter{figure}{-1}
\figcaption{(b). Same as Fig \protect\ref{fig:cdmsni}.(b) but for the $\Lambda $CDM model
\label{fig:lcdmsni}}
\end{center}
\end{figure}


\begin{figure}[htbp]
  \begin{center}
\leavevmode
\plotone{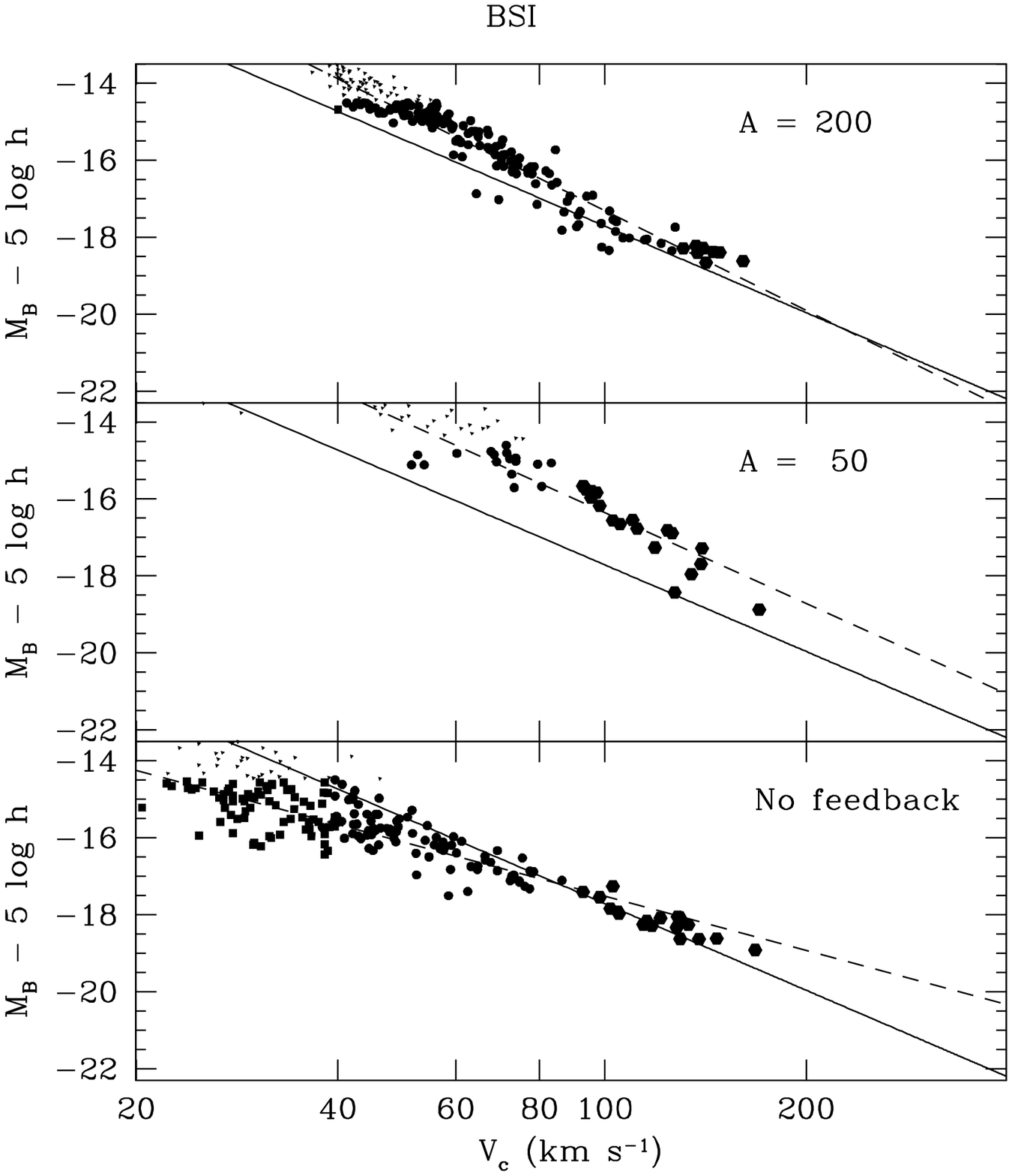}
\figcaption{(a). Same as Fig \protect\ref{fig:cdmsnb}.(a) but for  BSI model
\label{fig:bsisnb}}
\end{center}
\end{figure}


\begin{figure}[htbp]
  \begin{center}
\leavevmode
\plotone{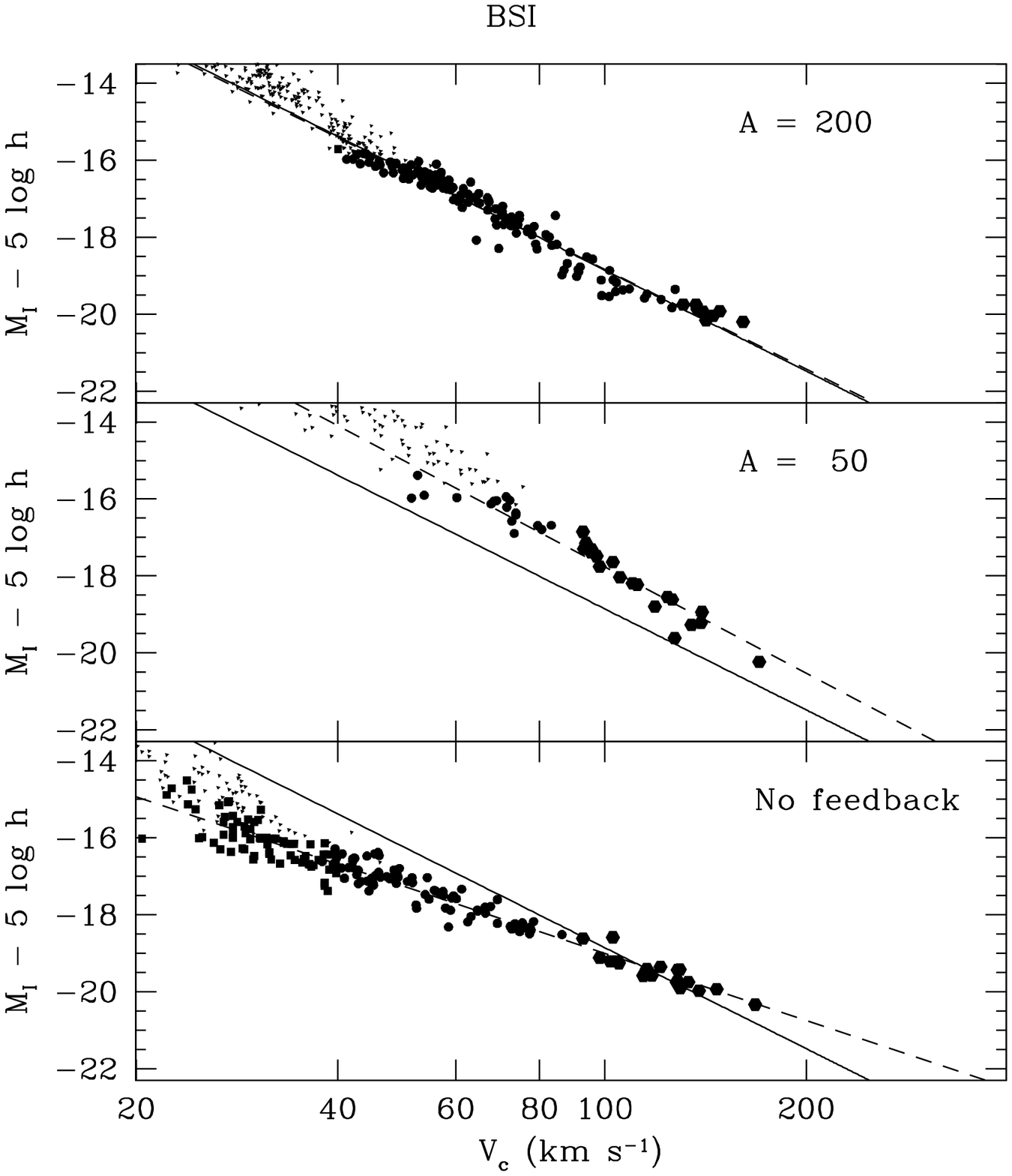}
\addtocounter{figure}{-1}
\figcaption{ Same as Fig \protect\ref{fig:cdmsni}.(b)  but for  BSI model.
\label{fig:bsisni}}
\end{center}
\end{figure}

\section{Fits to the TF relation}
\label{sec:fit}

From our galaxy samples in each scenario, we have computed magnitudes 
in the B, R, and I bands using the procedure of {\S} \ref{sub:gal}.
Figs.  \ref{fig:cdmsnb}, \ref{fig:lcdmsnb} and \ref{fig:bsisnb} are plots of the
$B$ and $I$ band magnitudes vs.  circular velocity for the simulated galaxies in
each of the three scenarios and for the three different feedback parameter values
considered:  $A=200$ (realizations where the evaporation of cold clouds is the
most important effect), $A=50$ (realizations in which supernovae are very
efficient at reheating the gas) and no supernova feedback (i.e.  $A=\beta=0$).
All galaxies are plotted, including those with $M_B>-16$.

To obtain the TF relation in each band considered, we have 
fit the relation between the log of the circular velocity 
$V_c$ and the magnitude $M$ in that band according to a formula of the general form
\begin{equation}
\label{eq:tfits}
M-5 \log h = a \log \left(\frac{V_c}{100\; {\rm km
\;s}^{-1}}\right) + b 
\end{equation} 
by linear least-squares regression. In order to come closer to the observational situation,
the fits include only ``bright'' galaxies ($M_B < -16$). 
 \begin{table*}
\caption{Parameters of linear fits ($a$, $b$) and  scatter ($\Delta
M$),
 for the magnitude-circular velocity relation in different bands
and models. Fits to the observational Tully-Fisher relation
(Pierce \& Tully 1992) are also shown. \label{tab:fits1}}
\small
\begin{tabular}{lcccccccccccc}
\tableline \tableline
   & & &  \mbox{$a$} & &&&   \mbox{$b$} & && &
\mbox{$\Delta$M} \\
   \cline{3-5} \cline{7-9} \cline{11-13}
Model &  Feedback Parameter & \mbox{$B$} & \mbox{$R$} & \mbox{$I$} &  &
\mbox{$B$} & \mbox{$R$} & \mbox{$I$}&  &\mbox{$B$} & \mbox{$R$} &
\mbox{$I$}   \\
\tableline

Observ. &----- &  -7.48 &  -8.23 & -8.72 & & -17.71 & -18.47 & -18.85 & &
0.34& 0.24& 0.24\\
\\
 &   A=200 & -6.08 &  -6.60 & -6.56 & & -16.21 & -17.35 & -16.24 && 0.46&
0.35&0.36 \\
CDM    &  A=50 &-7.47 &  -7.49 & -7.49 & & -15.87 & -17.04 & -17.63 && 0.29&
0.22&0.21 \\ 

    &  A=0 &-4.88 &  -5.47 & -5.65 & & -16.87 & -17.81 & -18.34 && 0.59&
0.45&0.42 \\ 
\\
 & A=200 & -7.28 & -7.12 & -7.14 & & -17.68 & -18.75 & -19.04 & &
0.41& 0.33&
0.30 \\

$\Lambda$CDM &  A=50 &-8.17 &  -8.18 & -8.18 & & -16.99 & -18.14 & -18.72 && 0.39&
0.27 & 0.25  \\ 

    &  A=0 &-5.05 &  -5.65 & -5.89 & & -17.99 & -18.92 & -19.46 && 0.48&
0.33 & 0.31  \\ 
\\

&  A=200 & -8.60 & -8.59 & -8.59 & & -17.31 & -18.29 & -18.84 & &0.37 &
0.28 & 0.27 \\

BSI  &  A=50 &-7.86 &  -8.90 & -9.20 & & -16.35 & -17.23 & -17.75 && 0.47&
0.36&0.33 \\ 

    &  A=0 &-4.67 &  -5.48 & -5.80 & & -17.52 & -18.44 & -18.99 && 0.45&
0.32&0.32 \\ 

   \tableline  \tableline
\end{tabular}
\end{table*}

Table \ref{tab:fits1} lists the resulting regression coefficients $a$ and
$b$ and the rms scatter $\Delta M$ expressed in magnitudes 
in the $B$, $R$, and $I$ spectral band for the 
observations (Pierce \& Tully, 1992) and for the simulated
CDM, $\Lambda$CDM, and BSI galaxy catalogs.  For each scenario, 
coefficients are computed for each of the supernova feedback 
parameters.  The linear fits
are also shown 
in Figs.  \ref{fig:cdmsnb}, \ref{fig:lcdmsnb} and \ref{fig:bsisnb}.

The results shown in the table and figures illustrate the sensitivity of the TF
relation to self-regulating processes affecting star formation in the halos.
The different cosmological scenarios correspond to varying characteristic
merging histories, while the different values of $A$ correspond to varying
supernova feedback.

Before discussing the issue of self-regulation, it is useful to point
out some general trends that can be derived from the fits:
Particularly noteworthy are the trends in the scatter and zero points.  All
models have their smallest scatter in the $I$ band, as do observed galaxies, and
in all three scenarios the $B$-band scatter is about 0.10 mag larger than the
$I$-band scatter, in agreement with the observed difference.  We also find the
same tendency of higher (more negative) zero points for the $I$ band than for
the $B$ band.  On the other hand, we do not find a systematic increase of the
TF slope toward longer wavelengths, at least when feedback from supernovae is
switched on. In all models, faint galaxies $(M_B>-16)$ seem to be systematically dimmer than
predicted by the TF fit for their rotational velocities.  This is consistent
with results from semi-analytic models (see e.g.  CAFNZ, HCFN), although in
these models, the departure from linearity is more pronounced for halos with
$V_c \leq 150$ km/s than in our simulations.

Concerning the effects of different cosmological scenarios, one sees that the
slope of the TF relation is steeper (i.e.  more negative) for less evolved
models ($\Lambda$CDM and BSI).  In the CDM model, galaxies with high circular
velocity have a lower luminosity than that predicted from the observed TF
relation.  An explanation for this trend is that in the CDM scenario, galaxies
tend to have large star formation rates at relatively early epochs.  As a
result, these galaxies have already transformed most of their gas into stars,
and the remaining gas does not cool fast enough to support a constant SFR,
because it is too hot and the density is too low.  The net result is a decline
in the SFR of these galaxies from $z\la 0.5$ up to the present epoch. A more
detailed treatment is given in Elizondo et al., 1998.

Table \ref{tab:fits1} and the figures 
show that the slope of the TF relation depends quite
strongly on the strength of supernova feedback. For low
values of $A$ (i.e. efficient gas reheating and less
evaporation of cold clouds), 
pressure gradients are effective in driving the gas away from the
center of halos with low circular velocity ($\leq 150 $ km/s).  
This results in a decrease in the star formation rate 
and hence in fainter galaxies. On  the other hand, halos with high
circular velocity are hardly affected by supernovae feedback, because
high-density gas efficiently radiates the energy released by
supernova explosions, keeping pressure gradients low. In this regard,
our simulations are in good agreement with previous work (Katz et al.
1992, Navarro and White 1994). 
Therefore, the TF relation  is steeper
(more negative slope) for simulations with lower $A$ values.
For $A=200$, the more evolved the scenario 
(CDM versus $\Lambda$CDM, and BSI), the
less steep the TF relation.  A possible explanation
is that in evolved scenarios, smaller halos tend to merge after having consumed
most of their gas, so that there is a shift toward higher
circular velocities without a corresponding increase in 
luminosity.

A very interesting result concerns the scatter of the TF relation in simulations
with $A=0$ (no heating or evaporation, but with metal enrichment).  $\Delta M$
is somewhat higher than for simulations with feedback.  Nevertheless, it is
still comparable with the observational estimates.  This seems to imply that the
origin of the small scatter in the TF relation cannot be attributed solely to
nonlinear feedback processes from supernova explosions.  We will discuss this
point in {\S} \ref{sec:discussion}.  The slope of the TF relation for $A=0$ is
quite similar in all three cosmological models.  It is flatter than in
simulations with feedback and very different from the observations.  However,
the luminosities of large circular velocity halos are not strongly affected by
turning off feedback.


\begin{figure}[t]
  \begin{center}
    \leavevmode
\plotone{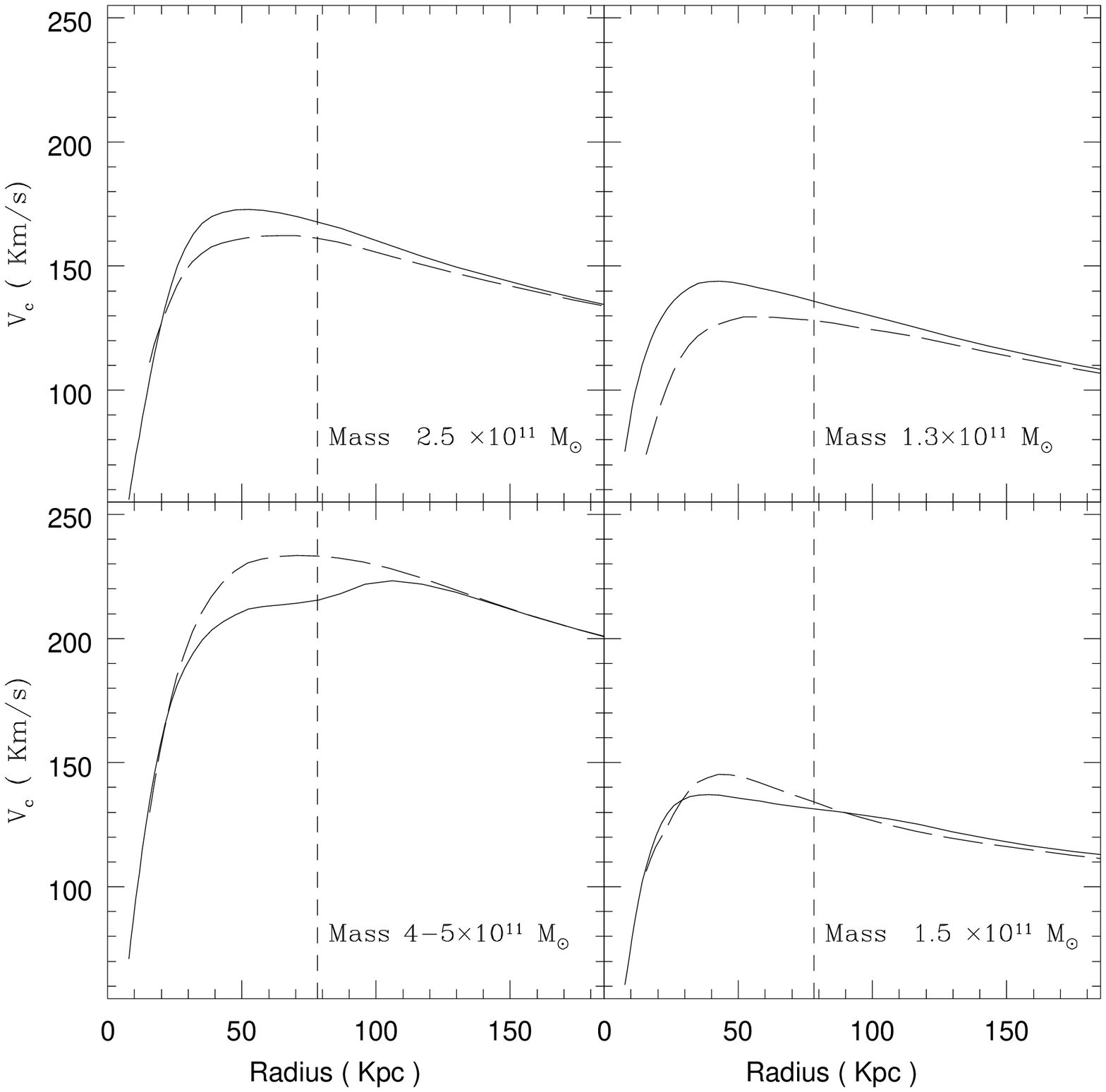}
    \figcaption[TULLY8.ps]{Effects of resolution on the
      circular velocity profile of 
       4 more massive  halos in one of the $\Lambda$CDM
  simulations reported      in      the paper. 
  The solid line is the circular velocity for the high-resolution simulation
  ($256^3$). Dashed lines correspond to the halos found in the same
  realization that was re-run at  lower resolution ($128^3$). 
The vertical dotted
 line represents the limiting radius for these halos according to our
 galaxy finding algorithm. 
    \label{fig:vres}}
  \end{center}
\end{figure}

\section{Numerical effects }
\label{sec:resolution}

Previous tests of our numerical simulation code were reported in YK$^3$.  For
the present study, several additional controls and tests were carried out to
quantify the effects of numerical resolution on our results.
As explained in {\S} \ref{subsec:procs}, one of the $\Lambda$CDM realizations
was simulated at two different resolutions.  We can thus study the
sensitivity of both the circular velocity estimate
and the magnitude assignment to numerical resolution and check for
biasing in the slope of the TF relations in different bands.


\begin{figure}[t]
  \begin{center}
    \leavevmode
\plottwo{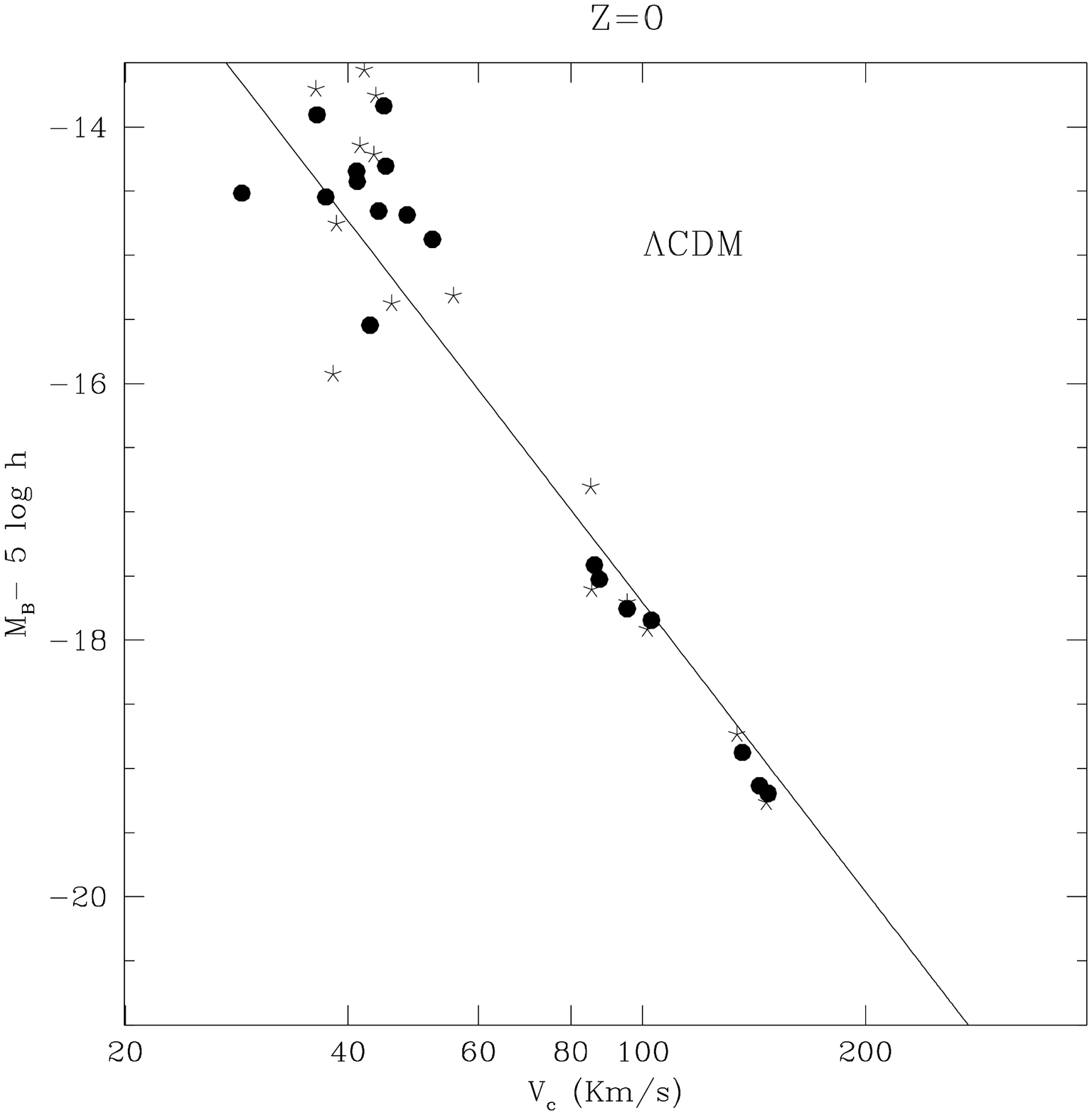}{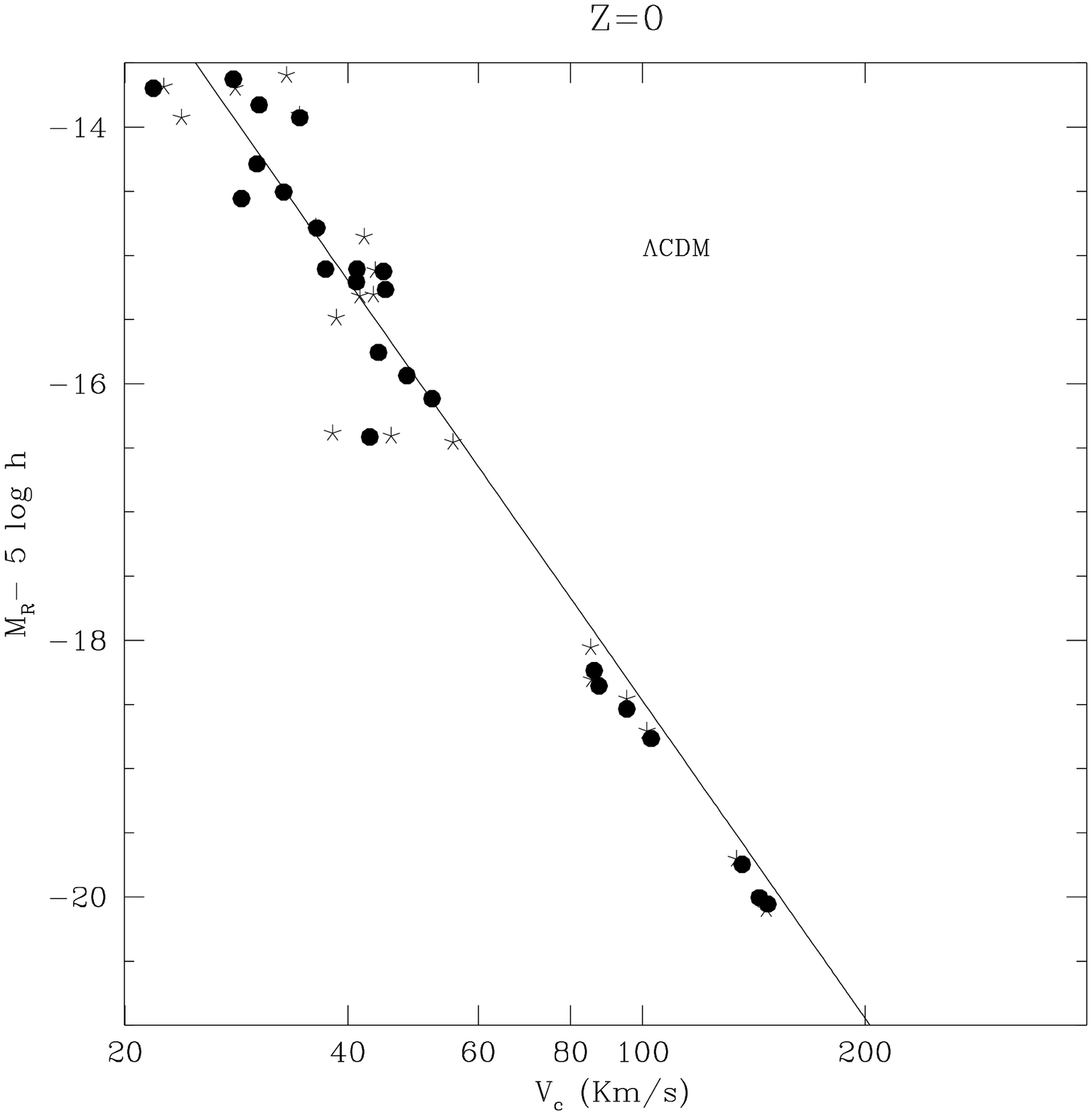}
    \figcaption[TULLY9a.ps,TULLY9b.ps.ps]{ Effects of resolution on the TF relation (in  $B$ and $R$ bands)
for one of the $\Lambda$CDM simulations reported in this paper.  The solid
dots represent galaxies found in the $256^3$ simulation (20 kpc
resolution). Star points  are galaxies found in the same simulation but
run  with $128^2$ cells (40 kpc resolution) and particles. Solid line
represents the observational fit to the TF relation (Pierce \& Tully 1992).
    \label{fig:tulyres}}
  \end{center}
\end{figure}

Fig.  \ref{fig:vres} shows the circular velocity
profiles as a function of radius for the four most massive halos found in the
test simulation. The circular velocity estimates at the
limiting radius for these halos differ by less than 10\% 
despite the factor of two difference in cell width.

In Fig.  \ref{fig:tulyres}, we plot magnitudes vs. circular velocities in
the $B$ and $R$ bands for the galaxies found in the simulation with $256^3$
particles and cells (19.5 kpc resolution, solid dots) and in the $128^3$
simulation (39 kpc resolution, asterisks).  
The effects of resolution on magnitude
are more pronounced for low circular velocity halos than for high circular
velocity halos. Nonetheless the linear
fit to the relation is hardly affected.  
In the high resolution simulation, we
find that $\Delta M_B$ is reduced by $\sim 20\%$ 
and $\Delta M_R$ by $\sim 30\%$
as compared with the results for the 
low-resolution simulation.


\begin{figure}[t]
  \begin{center}
    \leavevmode
\plotone{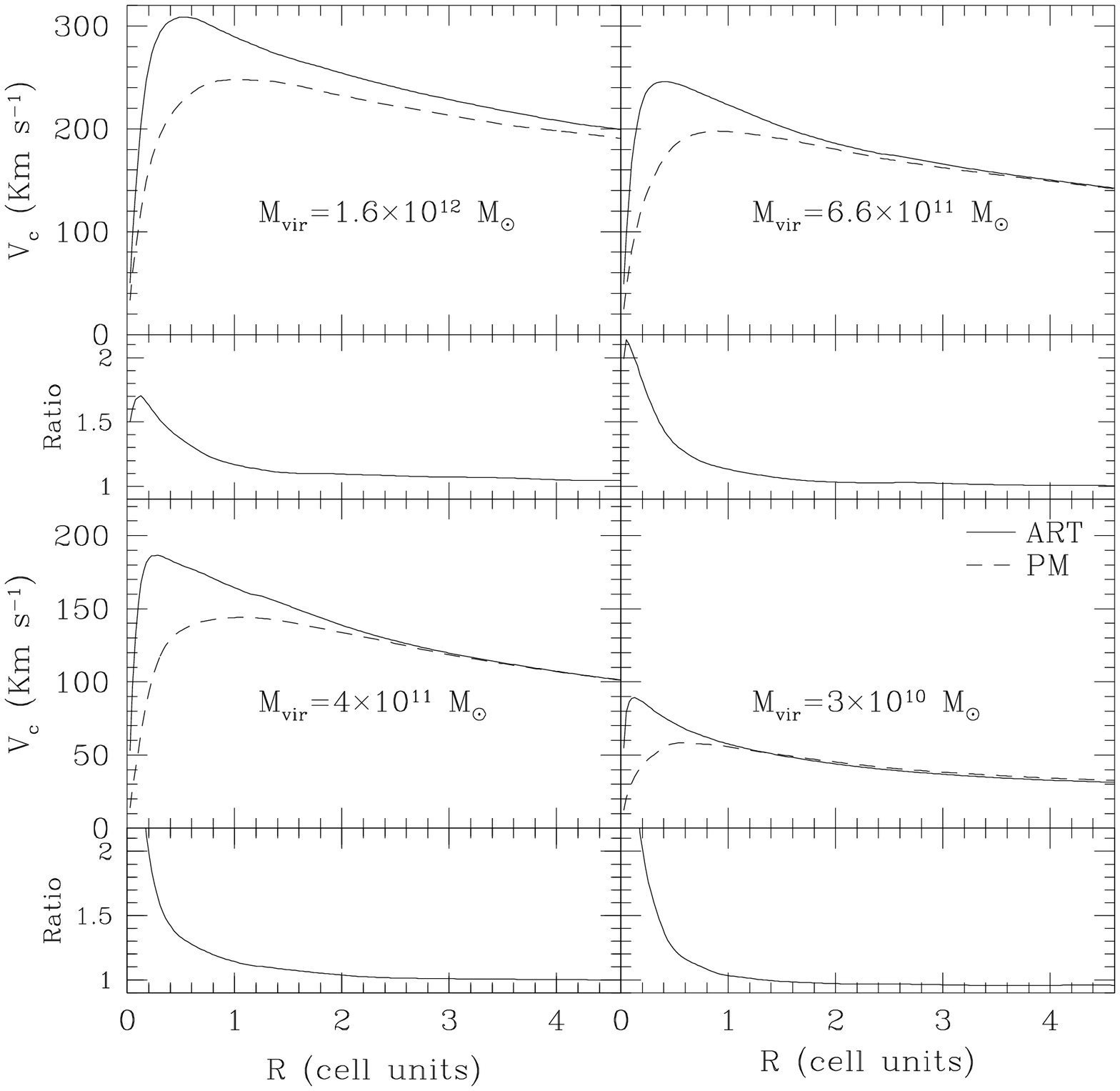}
    \figcaption[TULLY10.ps]{Circular velocity profiles 
      as a function of radius (cell units) for four halos found in 
      a CDM realization simulated using two 
      different N-body codes: a high resolution $(\sim0.2 kpc)$ ART code 
      (Kravtsov, Klypin \& Khokhlov 1997) (solid lines) and our PM 
      N-body code (dashed lines).
      Lower boxes in each panel  show the ratio of the two circular
      velocity estimates as a function of radius.
     \label{fig:andrey}}
  \end{center}
\end{figure}

From this analysis, we conclude that our star-gas model 
results in galaxies with properties that are robust with respect to
a factor of two change in numerical resolution.

As an independent test of the effects of resolution on the circular velocity
profiles of dark halos, one of the CDM realizations was also simulated using the
{\em Adaptive Refinement Tree (ART)} N-body code of Kravtsov, Klypin and
Khokhlov (1997).  The formal resolution of this simulation, at $z=0$ is $\sim
0.2 kpc$.  Despite the relatively low resolution of our simulation, the overall
structures of dark halos are similar to those of the ART simulation.  We have
identified the most massive dark halos and computed the velocity profiles, for
dark matter only, in both simulations.  In Fig \ref{fig:andrey} we show the
results of this comparison for four typical massive halos.  As expected, the
differences are striking at scales smaller than our cell resolution (39 kpc).
However, at the two-cell radius, our circular velocity estimate 
coincides almost exactly with the high-resolution estimate. 
The peak of the velocity curve of halos found in the ART simulation
is up to a factor of $\sim 2$ larger than the maximum of the velocity profile
from our simulation.  Nevertheless, circular velocities are assigned at
the two-cell radius, not at the peak, and luminosities are also defined
within the two-cell radius for massive galaxies.

\section{Luminosity functions}
\label{sec:luminosity}

Semianalytical models of galaxy formation have found it very difficult to
reconcile the observed zero-point of the TF relation with the
overall amplitude of the luminosity function (LF). It has been argued (Frenk
et al. 1996) that this problem is related to the overabundance of dark matter
halos predicted in all CDM cosmologies.  Dark matter simulations indicate that
the mass function of galactic halos has a slope of $\alpha \sim -2$ at the low-mass
end, which is steeper than any estimate of the observed field galaxy luminosity
function.  Hydrodynamical simulations such as those reported in this paper and
semianalytical models (i.e.  CAFNZ) have demonstrated that the faint end of the
luminosity function is very sensitive to the combined effects of mergers and
supernova feedback.


\begin{figure}[h]
  \begin{center}
\leavevmode
\plotone{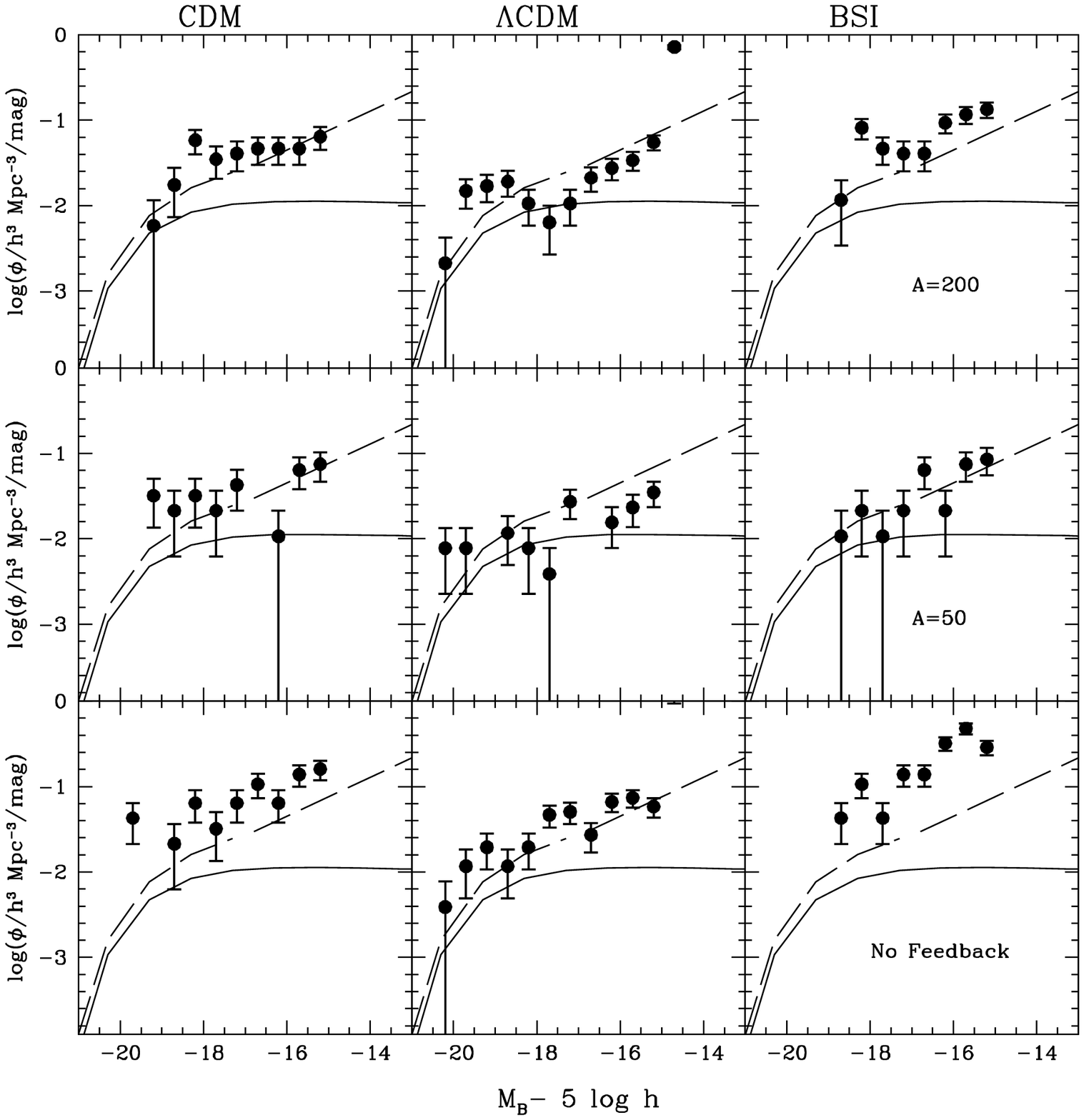}
\figcaption[TULLY11.ps]{Luminosity function (LF) in the B
  band. Filled circles represent the LF obtained from the numerical
  galaxies generated in   simulations with different feedback 
  parameters (top panels: $A=200$,  central panels: $A=50$, 
   and bottom panels: no feedback ($A=0$)). Error bars represent Poisson
   errors in each magnitude bin. 
   Solid lines 
  are the best-fitting Schechter function from the Stromlo-APM survey
  (Loveday et al. 1992).
  Dashed line  correspond to the best fit  to the LF obtained from ESP
 galaxy redshift survey (Zucca et  al. 1997) \label{fig:funlumb}}.
\end{center}
\end{figure}

\begin{figure}[t]
  \begin{center}
\leavevmode
\plotone{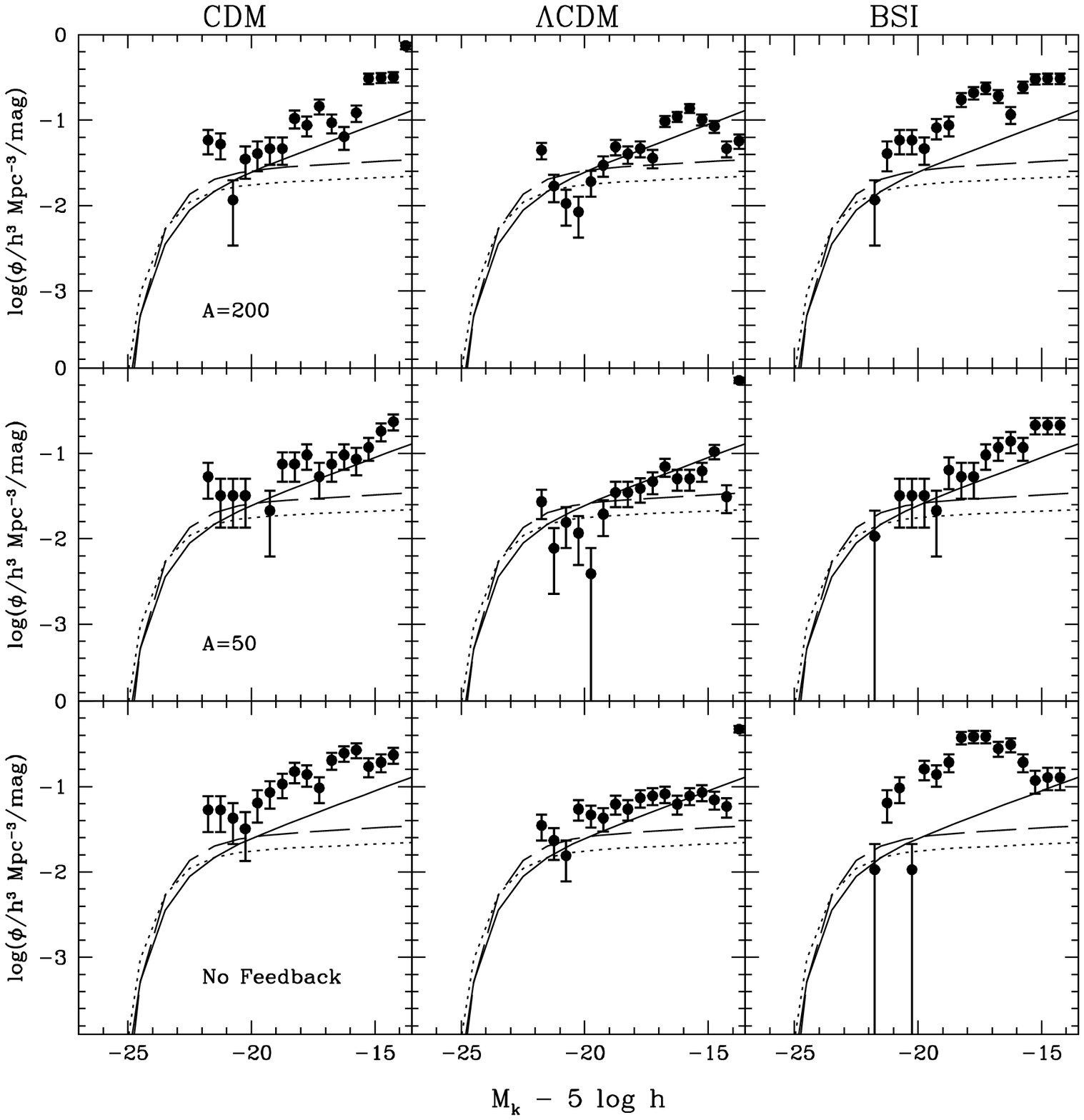}
\figcaption[TULLY12.ps]{Same as in figure
  \protect\ref{fig:funlumb} but for the $K$ band. Solid line is the
  best-fitting Schechter function from Szokoly et al. (1998). Dotted
  line is the best fit to the LF obtained by Gardner et al. (1997).
 Dashed line corresponds to the best fitting LF obtained by
 Glazebrook et al. (1995) \label{fig:funlumk}}. 
\end{center}
\end{figure}

In Figs.  \ref{fig:funlumb} and
\ref{fig:funlumk} we show the luminosity functions (solid circles) in the $B$
and $K$ band estimated from the numerical galaxies generated in the realizations
of the three cosmological simulations with different feedback parameters (top
panels:  $A=200$, central panels:  $A=50$ and bottom panels:  no feedback).
Observational estimates for the local luminosity functions are also shown.  For
the $B$ band, the solid line represents the Schechter function with $\alpha \sim
-1$, which best fits the observational data from the Stromlo-APM survey (Loveday
et al.  1992).  The dashed line shows a more recent estimate of the LF found
from the ESO Slice Project (ESP) galaxy redshift survey by Zucca et al.  (1997).
These authors find a Schechter function with $\alpha=-1.22$ as an acceptable
representation of the LF over the entire range of magnitudes $M_B - 5 \log h
\leq -12.2$.  However, their data also suggest the presence of a steepening of
the luminosity function for $M_B -5 \log h \geq -16.8$.  Such a steepening at
the faint end of the LF is well fitted by a power law with slope $\sim -1.6$.
The range of magnitudes of the galaxies in our simulations ($-20 \la M_B \la
-15$) allows us to study the faint-end slope of the LF only.


In the $K$ band plots, we show the Schechter functions fitting the most
recent observational data in the near infrared band.  Solid lines represent the
best fit to the Szokoly et al.  (1998) survey ($\alpha = -1.27$).  Dotted lines
are the best fit to the LF obtained by Gardner et al.  (1997) ($\alpha =
-1.03$).  Dashed line corresponds to the best-fit LF obtained by Glazebrook et
al.  (1995) ($\alpha = -1.04$).  

The lack of bright galaxies, due to the small simulated box, precludes drawing
conclusions about the observed break at the bright end of the LF.  The faint-end
slopes of the $B$-band LF derived from our data are steeper ($\alpha \sim -1.5$
to $-1.9$, for $-18 \leq M_B \leq -15$) than the estimate of Loveday et
al.  (1992), but they are in rough agreement with the recent estimates from
the ESP survey.

For the $K$-band, our galaxies have magnitudes much fainter ($-22 \la M_K \la
-14$) than those of the near-infrared redshift surveys ($M_K \la -21$).
Therefore, our data should be compared with the extrapolation of the
best-fitting Schechter functions towards fainter magnitudes.  Our data show a
clear trend toward a steepening of the slope of the luminosity function at the
faint end, in agreement with the recent estimates of Szokoly et al.  (1998), as
compared with the results of Glazebrook et al.  (1995) and Gardner et al.
(1997).  The slopes for the faint end of the LF from our data are in the range
$\alpha \sim -1.2$ to $-1.5$.

In view of the dependence of the luminosity function normalization on
merging rates and supernova feedback we can explain the higher faint-end BSI 
LF as follows: In the BSI simulations,
galaxies tend to form later than in the $\Lambda$CDM
or CDM simulations. When supernova feedback is included, it
most strongly affects star formation  in  small galactic halos, which end up
fainter, as noted above. 
The net result is a lower normalization of the luminosity
functions for BSI with supernova feedback than without it. 
Efficient reheating of gas from supernovae (low $A$) results in a
luminosity function with lower normalization than with more evaporation 
(high $A$). 
Again this effect is more
pronounced in the BSI simulations, because most of the galactic
halos have small potential wells,  so gas can more easily escape due
to the relative large pressure gradients generated by supernova explosions.

\section{Discussion and Conclusions}\label{sec:discussion}

Using hydrodynamical simulations of galaxy formation with star formation and
supernova feedback we have found that it is possible to reproduce the slope,
offset, and scatter of the TF - relation.  It is quite intriguing to ask which
elements of the approach are essential for the different parts of the relation:
In simulations without supernova feedback (energy input and mass transfer from
supernova is switched off, but metal enrichment is left on), the slope of the
resulting TF - relation is too shallow $a \sim -5$, or $L \propto V_c^2$, with
nearly equal zero-point in all three scenarios (cf.  Table \ref{tab:fits1}).
This slope agrees with estimates from the Press-Schechter theory (Cole and
Kaiser, 1989) which incorporate cooling but ignore feedback and assume
that all cold gas is converted into stars.

According to our results, the slope of the TF relation depends quite strongly on
the level of feedback.  The effects of supernova feedback are certainly much
less significant for galaxies with $V_c \ga 150$ km s$^{-1}$ (cf.  Figs
\ref{fig:cdmsnb}--\ref{fig:bsisni}) in fairly good agreement with previous
results (e.g.  Navarro \& White 1993; Katz et al. 1996). 
Supernova feedback tends to have a stronger effect on  star formation 
in low circular velocity halos.  Hence supernova feedback causes a steepening of
the slope, because low-mass halos become fainter while the luminosity of massive
halos remains basically unchanged. 
This feature of the TF relation is the common to the three
cosmological scenarios considered here.

The slope of the TF is the same for all photometric bands when feedback is
considered, while for simulations without feedback,  the TF-slope is
steeper for redder wavelengths. 
These results suggest that the
observed passband dependence of slope of the TF relation should not be
attributed to different star-formation histories of galaxies. 
Although corrections for dust extinction have been made in the analysis of observations 
(Pierce and Tully, 1992), these corrections are rather difficult to estimate and hence uncertain. 
Dust extinction 
could well be responsible for the differences between  the B-band  and I-band 
 TF slopes.  
Moreover, recent semianalytical models which include dust extinction
also find  a varying slope of the TF with wavelength 
(Somerville \& Primack 1998;  Firmani \& Avila-Reese 1998).  
An important extension of the present framework will be to incorporate
the effects of dust extinction including a refined model for the metal production
of stars. 

Another important ingredient of the TF relation is the intrinsic
scatter.   As we have already mentioned, a long standing question 
concerns whether this scatter is a consequence of cosmological conditions, or 
whether it is rather produced by the nonlinear physical processes associated with
star formation that occur in the interior of halos.
Interestingly, the TF scatter does not change much   
 as the supernova feedback parameter is varied, as long as feedback is
 not turned off completely (all SN energy radiated away immediately). 
However, the scatter 
in simulations with feedback turned off is about double the observed TF scatter. 
Feedback processes tend to reduce this scatter, 
because  they  are  coupled to 
merging rates and dynamical evolution.  For instance, the rms
scatter in BSI simulations is hardly affected by feedback,  while
in CDM simulations, the rms scatter with feedback is half that of 
simulations without feedback.
Hence, the idea of Eisenstein \& Loeb (1996)  that
small  TF scatter is related to nonlinear feedback mechanisms could
indeed be correct, but some nonlinear element in addition
to supernova feedback must play an essential role in
reducing it.

In a recent paper, Somerville \&
Primack (1998)  have carefully compared different semianalytical
model predictions for the TF scatter and slope. 
They conclude that the predictions are sensitive to different assumptions concerning supernovae 
feedback. In particular, the ``Durham semianalytical model'' (CAFNZ) 
has  a feedback model that is too strong, resulting in a nonlinear TF
relation at low circular velocities. The ``Munich  model'' (KWG) has a
milder feedback model and consequently, it gets a flatter slope for the
TF, in agreement with our numerical results.
Avila-Reese, Firmani and
Hern{\'a}ndez (1998) use a different  semianalytical approach for
the virialization of baryonic material  in rather  
isolated (mass accretion dominates over  major mergers) 
dark matter halos, without gas outflow from the disk. 
They  find a scatter in the TF relation which is somewhat higher (within
a factor of 2)  than in observations (Pierce \& Tully 1992),
in  fairly agreement with  what we find in our simulations
with no feedback. In these semianalytical models, the 
scatter in the TF relation can be traced back to the scatter in 
the mass accretion histories of halos of similar mass (e.g. Avila-Reese
1998).


\begin{figure}[t]
  \begin{center}
    \leavevmode
\plotone{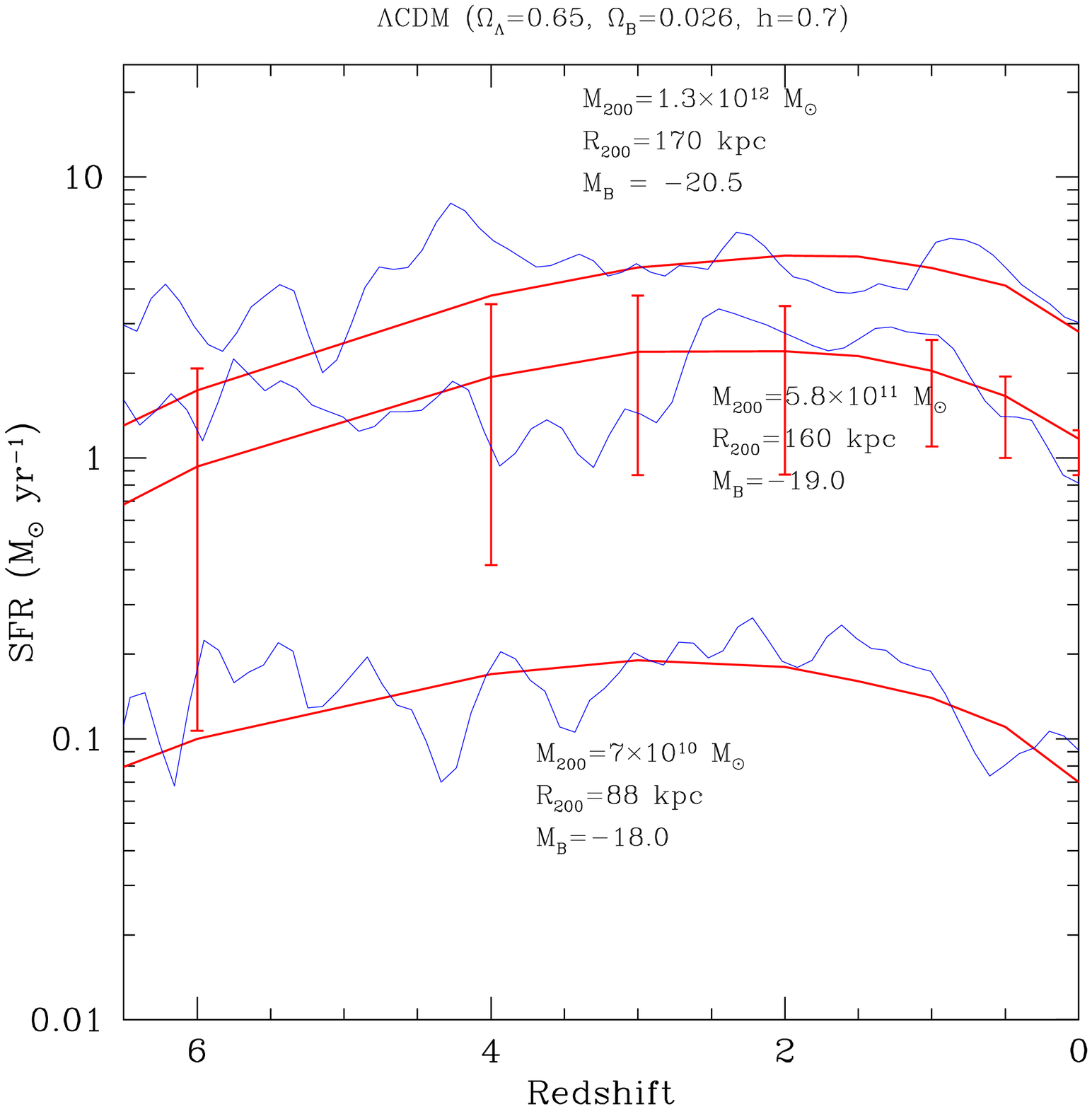}
    \figcaption[TULLY13.ps]{Comparison of the redshift evolution
      of the Star Formation Rate for three different halos in one of our 
      $\Lambda$CDM simulation (thin solid lines), 
      with averaged predictions from semianalytical  galaxy formation
      models (Firmani \& Avila-Reese 1998). The error bars correspond
      to the maximum and minimum values of the SFR  in the
      Montecarlo realizations of mass accretion histories for the
      corresponding halo.}
    \label{fig:vladimir}
  \end{center}
\end{figure} 

In this regard, it is interesting to compare the redshift evolution of the star
formation rate for some of our halos with predictions of semianalytical
calculations, kindly provided by V.  Avila-Reese.  In Fig \ref{fig:vladimir} we
show the star formation rate (SFR) as a function of redshift for three halos of
different mass in one of our $\Lambda $CDM simulations.  The thick solid lines
represent the average SFR for halos of same virial mass, computed by means of the
semianalytical model developed by Firmani and Avila-Reese (1998).  The error
bars represent the maximum and minimum SFR values at different redshifts for the
2000 Monte Carlo realizations of mass accretion histories for this halo.  This
semianalytical model of formation of disk galaxies includes a simplified
hydrodynamical treatment of the gas and models the effects of supernovae on the
cold gas clouds as an increase in their turbulent motion.  This model differs
from our picture of cloud evaporation, but it nonetheless provides an effective
feedback mechanism.  The scatter in the TF relation for these models is somewhat
higher than the one we obtain in the simulations.  However, the overall
agreement between these semianalytical calculations and our simulations is
remarkable.  (See Firmani \& Avila-Reese 1998 for further details).

The answer to the question of whether the scatter of TF is a
consequence of cosmological conditions, or rather results from 
nonlinear processes associated with star formation in the dark halos,
could well be: {\em  The scatter of the TF is due to 
a combination of effects in which nonlinearities of 
several kinds play essential roles.}  
The  intrinsic scatter that results in the formation and
virialization of galactic halos is reduced  when  
additional nonlinear processes due to the supernovae feedback loop are taken
into account. Those semianalytical models with a highly nonlinear
model of the baryonic physics tend to find small scatter. As
a general rule, numerical simulations that self-consistently 
incorporate gravity, hydrodynamics and feedback effects will include more nonlinearity 
than simpler approximations. Hence, it is
not surprising that a lower scatter in the TF relation  
is found in simulations than in semianalytical model
predictions.


We have also estimated the luminosity functions in $B$  and $K$ bands from
the simulated galaxies.  The relatively small volumes of
our present simulations preclude a reliable estimate of the
full LF in the different cosmological scenarios, since as explained
earlier bright galaxies are undersampled or completely absent due to the 
lack of high-density regions. With this caveat in mind, we have compared
the number density of faint objects with
observations. The faint end of the LF is very much affected by
supernova feedback. This is especially important for the less
evolved BSI simulations, in which a high number density of faint
objects is found. In all simulations, we systematically find steeper
slopes for the $B$ band luminosity functions than the estimates from  the
Stromlo-APM survey (Loveday et al. 1992), but they are compatible
with recent estimates from the ESO Slice Project (Zucca et al. 1997). 

Despite possible uncertainties in the
determination of the LF from our simulations, we can conclude that,
with an adequately chosen value for the feedback parameter,  our
model  of self-regulating star formation can reproduce the correct 
TF zero point and the right number density of objects, at  least for the
$\Lambda$CDM model. For the CDM simulations,  halos tend to have a
lower luminosities for a given circular velocity, so the zero point of
the TF relation is incompatible with observations, while for the BSI
model, the low normalization of the spectrum produces far too many
faint objects. The number of objects can be reduced by assuming a  small
value for the feedback parameter ($A=50$), which means a large thermal
reheating. In this case, the  galaxies are fainter and the TF
zero point is not compatible with the observed one. 

We saw in Section \ref{sec:resolution} that our results are fairly 
robust with respect to changes in numerical resolution and details of
the galaxy finding scheme.  Doubling the resolution did not significantly
affect the TF relation, although some individual galaxy colors were
shifted.  As resolution is further increased, we expect that some adjustment
of parameters in our modeling of processes occurring below the
limits of resolution will be required. 

A final conclusion of this paper is that self-regulation of star formation such
as supernova feedback and hyrodynamical processes must play an essential role in
explaining the observational properties of galaxies.

\acknowledgements
We wish like to thank  Claudio Firmani and Vladimir Avila-Reese for
helpful discussions and for providing us with comparisons from their
semianalytical calculations. We would also like to thank Andrey
Kravtsov for performing one of our CDM realizations using  the ART code and 
computing the velocity profiles. 
This work  has been partially supported by the DGICyT (Spain) under
project number PB93-0252. R.K. gratefully acknowledges a fellowship of the
DFG (Germany) during part of this work.


\end{document}